\documentclass[aps,prb,floatfix,twocolumn,groupadresses,showpacs]{revtex4}
\usepackage{graphicx}
\usepackage{amsmath}
\usepackage{amssymb}

\begin{document}

\title[Structural deformations of two-dimensional planar structures] 
{Structural deformations of two-dimensional planar structures 
under uniaxial strain: The case of graphene}

\author{Zacharias G. Fthenakis}
\affiliation{Institute of Electronic Structure and Laser, FORTH, Heraklion, Greece}

\author{Nektarios N. Lathiotakis}
\affiliation{Theoretical and Physical Chemistry Institute, National Hellenic
Research Foundation, Vass. Constantinou 48, GR-11635 Athens, Greece  }
\date{\today}

\begin{abstract}
In the present work, a method for the study of the structural deformations
of two dimensional planar structures under uniaxial strain is presented.
The method is based on molecular mechanics using the original stick and
spiral model and a modified one which includes second nearest neighbor
interactions for bond stretching. As we show, the method allows an accurate
prediction of the structural deformations of any two dimensional planar
structure as a function of strain, along any strain direction in the
elastic regime, if structural deformations are known along specific strain
directions, which are used to calculate the stick and spiral model
parameters. Our method can be generalized including other strain conditions
and not only uniaxial strain. We apply this method to graphene and we test
its validity, using results obtained from {\it ab initio}
Density Functional Theory calculations. What we find is that the original
stick and spiral model is not appropriate to describe accurately the
structural deformations of graphene in the elastic regime. However, the introduction of second nearest neighbor interactions provides a very
accurate description.
\end{abstract}
\pacs{61.48.Gh, 62.20.-x, 62.20.de, 62.20.dj, 62.20.dq, 62.20.F-, 62.23.Kn, 
62.25.-g}
\maketitle

%
%
%
%
%

\section{Introduction}
Undoubtedly, graphene is one of the most studied
materials in recent years. This is due to its exotic properties, like for
instance its high carrier mobility \cite{graphene_electron_mobility} and 
high thermal conductivity \cite{DT217, ghosh, susp_graphene_exper_TC} at
room temperature, its high strength \cite{PCCP_Fthenakis, strength}, etc,
which makes graphene one of the most interesting materials for future
nanoelectronic and nanomechanic applications. Following graphene, several two
dimensional (2D) materials have also gained interest, exhibiting
interesting mechanical \cite{mech_properties,mech_BN,mech_C2F_BN, mech_MoS2, mech_silicene_1,
mech_silicene_2,mech_phosphorene, Fthenakis_Si2BN} and electronic properties. 
The world of  2D materials that have been brought to the center of attention
recently \cite{review_2D_1, review_2D_2} includes several transition metal dichalcogenides
\cite{TMDs, mech_MoS2}, (like for instance MoS$_2$ or WS$_2$),  
hexagonal BN (h-BN) \cite{mech_BN,mech_C2F_BN, hex_BN,many}, 
Si$_2$BN \cite{Fthenakis_Si2BN, Madhu_Si2BN}, Si$_n$B$_m$ \cite{SiB, Si_nB_m,BSi3}, SiX and XSi$_3$ (X=B, C, N, Al, P) \cite{SiX},
CdS \cite{CdS_planar},
AlN \cite{hex_AlN,hex_AlN_theory,many,hex_AlN_mechanical}, SiC, InN and
GaN \cite{many}, C$_2$F \cite{mech_C2F_BN}, Silicene \cite{mech_silicene_1, mech_silicene_2, silicene,
silicene_germanene}, Germanene
\cite{silicene_germanene}, Siligene (SiGe) \cite{SiGe_2D}, Phosphorene
\cite{mech_phosphorene, phosphorene}, as well as 
several graphene allotropes, like pentaheptites and octagraphene
\cite{PCCP_Fthenakis,enyashin}, or other Carbon 2D allotropes, like
pentagraphene \cite{pentagraphene}, graphyne, graphydine 
\cite{enyashin,allotrop}, or graphene-based derivatives, like graphane
and graphone \cite{graphane, allotrop} etc.

A special class of these materials are those which are entirely planar,
like for instance several graphene allotropes (pentaheptites, octagraphene,
etc) \cite{PCCP_Fthenakis, enyashin, haekel}, as well as h-BN \cite{hex_BN}, 
Si$_2$BN \cite{Fthenakis_Si2BN, Madhu_Si2BN}, AlN, SiC, 
Si$_n$B$_m$ \cite{SiB,Si_nB_m,BSi3}, CdS \cite{CdS_planar}, 
XSi$_3$ with X=B,C,Al \cite{SiX} etc.
In this work, we present a method for the study of the mechanical response,
of these materials, e.g. bond stretching and angle bending deformations, 
in the presence of uniaxial tensile strain, providing analytic expressions
for these deformations along any strain direction. Our
method can be generalized including any other strain 
condition (i.e. not only uniaxial strain) and
is based on molecular mechanics assuming two different versions of
the so called \emph{stick and spiral model} \cite{stick_spiral}, which
has been employed previously for the study of the mechanical properties of
Carbon nanotubes \cite{Geng, Chang, Zhao1, Zhao2, Zhao3,Jiang}.

As an example, we apply our method to graphene, providing analytic
expressions for bond length and bond angle deformations 
under tensile strain. We test the accuracy of these expressions
using results we obtain from ab-initio density functional theory (DFT)
calculations. In particular, we calculate the structural deformations
of graphene under
tensile strain along the high symmetry arm chair and zig-zag directions,
as well as two other randomly selected directions, which are perpendicular
to each other. According to our findings,
the original stick and spiral model is not sufficient to provide an
accurate description of the mechanical deformations of graphene under
tensile strain in the elastic regime, since the DFT results can not be
reproduced accurately by the analytic expressions provided by that model. However,
due to the coupling between the bond stretching and angle bending terms,
which is inherently included in the modified stick and spiral model, this 
modified model provides a quite accurate description. Moreover, fitting
these analytic expressions to the DFT results we calculate the force
constants for bond stretching and angle bond bending for graphene, thus
allowing the prediction of the mechanical response of graphene in the elastic regime
for strain on any direction. 

\section{The deformation energy}
In molecular mechanics approach the deformation energy $U$ is a sum of
energy contributions from different deformation modes \cite{stick_spiral}.
In particular, $U$ is written as 
\begin{equation}
U=U_s+U_b+U_\omega+U_\tau+U_{vdw}+U_e,
\end{equation} 
where $U_s$, $U_b$, $U_\omega$, $U_\tau$, $U_{vdw}$ and $U_e$ correspond
to the energy contributions from bond stretching, bond angle bending,
bond inversion, bond angle torsion, Van der Walls interactions and
electrostatic interactions, respectively. Since tensile strain in a 2D
planar structure is in-plane strain, the terms $U_\omega$ and $U_\tau$
vanish. Moreover, since there are no interactions between different sheets
of those 2D structures, the terms $U_{vdw}$ and $U_e$ also vanish. Thus,
the deformation energy becomes
\begin{equation}
U=U_s+U_b.
\end{equation}
$U_s$ and $U_b$ may be expressed in several different ways 
(see for instance Refs.~\onlinecite{Davydov2010, kalosakas,lobo}).
However, the simplest way is to be expressed as a sum of harmonic
terms constituting the so-called \emph{stick and spiral model}.

According to the stick and spiral model, the deformation energy per unit
cell is written as a sum of energy contributions from each bond length
and bond angle deformation. Each of these contributions has a quadratic
dependence on the corresponding deformation, i.e. it is either of the
form $(1/2)k_s\delta l^2$ (for bond stretching), or 
$(1/2)k_b\delta \phi_{ij}^2$ (for bond-angle bending), where $k_s$ and
$k_b$ are the corresponding force constants, and $\delta l$ and 
$\delta \phi$ the bond length and bond-angle deformations for each
specific bond and bond angle, respectively. Thus, the deformation energy
per unit cell is
\begin{equation}\label{energy1}
U = \frac{1}{2}\sum_i \left(k_{s,i}\delta l_i^2 + \frac{1}{2}\sum_j k_{b,ij}\delta\phi_{ij}^2\right),
\end{equation}
where $i$ counts all the bonds inside the unit cell and $j$ counts the
bonds which form bond angles with bond $i$. The $1/2$ factor of the
second sum is to avoid double counting of the bonds.

In the description provided by the stick and spiral
model, bond stretching and bond angle bending are not coupled. 
The energy provided by Eq.~(\ref{energy1}) does not have any 
terms mixing these deformations. In addition, as we will see later,
in the minimization of the deformation energy under constant strain 
these deformations remain decoupled. More specifically, one arrives at
two independent systems of analytic equations one for stretching and
one for bending.  A more
accurate description would include a coupling term between these deformations. 
This can be achieved by introducing extra terms
describing the stretching of second nearest neighbor interatomic distances.
In the present work, we study both cases.

\begin{figure*}
\begin{center}
\includegraphics [width=0.95\textwidth,clip]{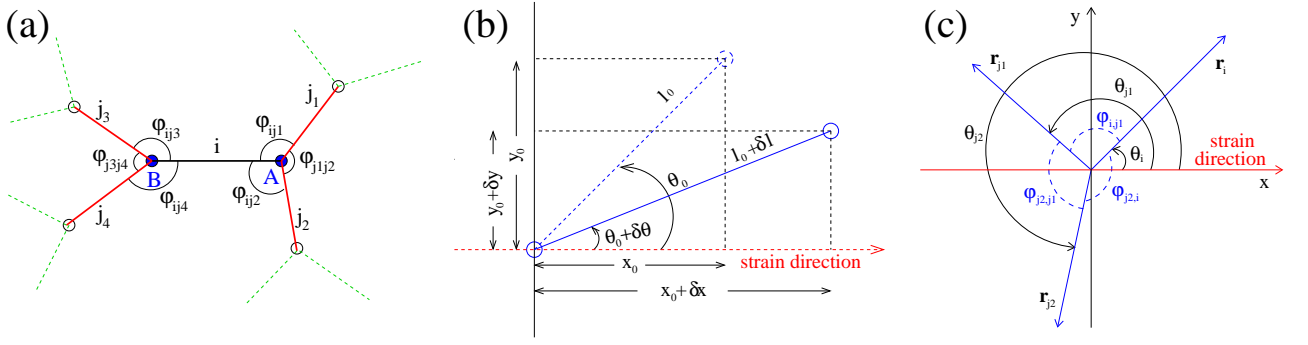}
\end{center}
\caption{\label{Fig1}
(Color online)
(a) Bond $i$ of atoms A and B. Atom A forms the bonds i, j1 and j2 with
its neighboring atoms and atom B forms the bonds i, j3 and j4,
(b) Bond and angle deformations under uniaxial strain, 
(c) Relation between $\theta_i$ and $\phi_{ij}$.
}
\end{figure*}

For a planar structure with three-fold coordinated atoms, there are 
three bonds and three bond angles per atom (see Fig.~\ref{Fig1}(a)). If
we label $i$, $j_1$ and $j_2$ the bonds of atom A and $i$, $j_3$ and $j_4$ those
of atom B, (the two atoms share the bond $i$), then the index $j$ of 
Eq.~(\ref{energy1})
takes the values $j_1$, $j_2$, $j_3$ and $j_4$. Moreover, since the
structure is planar, and all atoms remain in the plane under tensile strain
\begin{equation}
\phi_{ij_1}+\phi_{ij_2}+\phi_{j_1j_2}=\phi_{ij_3}+\phi_{ij_4}+\phi_{j_3j_4}=2\pi,
\end{equation}
where $\phi_{ij_1}$, $\phi_{ij_2}$, $\phi_{j_1j_2}$ are the bond angles
of atom A and $\phi_{ij_3}$, $\phi_{ij_4}$, $\phi_{j_3j_4}$ the bond
angles of atom B. Consequently,
\begin{equation}\label{delta_phi}
\delta\phi_{ij_1}+\delta\phi_{ij_2}+\delta\phi_{j_1j_2}=\delta\phi_{ij_3}+\delta\phi_{ij_4}+\delta\phi_{j_3j_4}=0.
\end{equation}
In the present work we study structures with only 3-fold 
coordinated atoms, since this is the most common case. However, the
generalization of our method to structures
with $n$-fold coordinated atoms, with $n\ne 3$, is obvious.

\begin{figure}[!tb]
\begin{center}
\includegraphics[width=0.475\textwidth,clip]{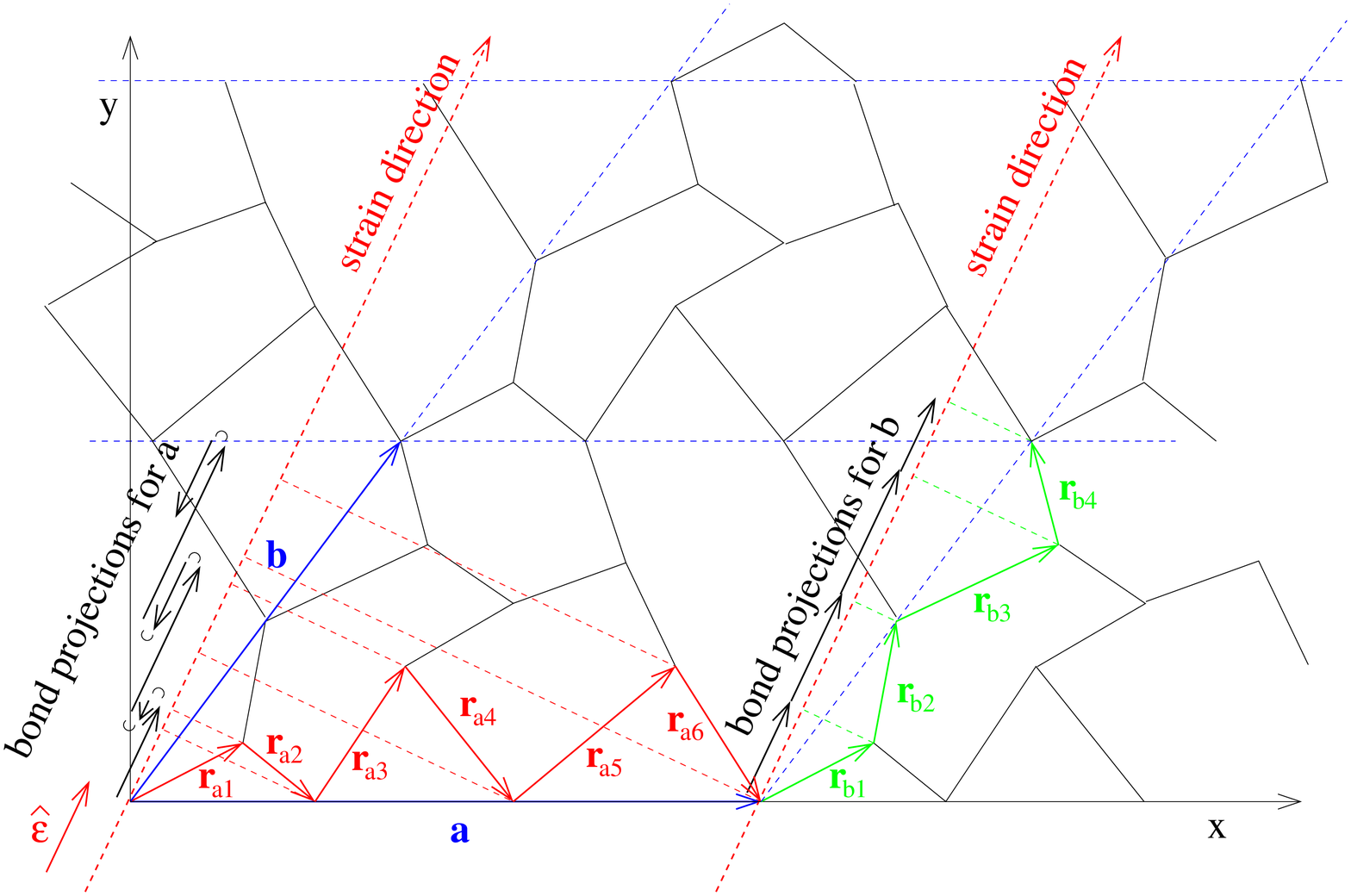}
\end{center}
\caption{(Color online) 
Periodic planar structure with 3-fold coordinated atoms strained
along the strain direction $\hat{\boldsymbol{\varepsilon}}$ (colored in red).
The unit cell vectors (colored in blue) are $\mathbf{a}$ and $\mathbf{b}$.
The vector sum of the vectors $\mathbf{r}_{ai}$ ($\mathbf{r}_{bi}$) 
corresponding to the red (green) colored bonds, constitute the unit cell
vector $\mathbf{a}$ ($\mathbf{b}$). The projection of those bond vectors
along the strain direction are shown with black arrows along the strain
direction.}
\label{projections}
\end{figure}

Due to symmetry reasons (if any), several bonds length deformations (as
well as bond angle deformations) may be equivalent with each other under
specific strain conditions. In that case, $U$ can be written as a function
of only the independent bond length and bond angle deformations per unit
cell, and Eq.~(\ref{energy1}) can be rewritten as
\begin{equation}\label{energy2}
U = \frac{1}{2}\left(\sum_i n_i k_{s,i} \delta l_i^2 + \frac{1}{2}\sum_i\sum_j m_{ij} k_{b,ij}\delta\phi_{ij}^2\right),
\end{equation}
where $n_i$ is the number of equivalent bond length deformations of type
$i$ and $m_{ij}$ the number of equivalent bond angle deformations formed
by the bonds which have independent bond length deformations of type $i$
and $j$. $i$ runs over the independent bond deformations only.
 
Under uniaxial strain, 
the deformation energy and the corresponding deformations $\delta l_i$ and
$\delta \phi_{ij}$ at the strained equilibrium can be found from
the minimization of the deformation energy subject to constrains describing
the strain condition. These constraints can be incorporated 
using the Lagrange multipliers technique. 
For constant uniaxial tensile strain $\varepsilon$ there is only one constraint
described by $\varepsilon = \delta L/L_0$, where $L_0$ is a length
along the strain direction and $\delta L$ the elongation of $L_0$ upon
that strain, which should be expressed as a function of the independent
variables $\delta l_i$ and $\delta \phi_{ij}$. Thus, the function which
should be minimized becomes
\begin{equation}\label{Lagrange}
\Lambda = U+\lambda(\varepsilon - \delta L/L_0),
\end{equation}
with $\lambda$ the corresponding Lagrange multiplier. 
Obviously, for different strain conditions, different
constrains will apply, which can be incorporated in Eq.~(\ref{Lagrange})
using the corresponding Lagrange multipliers.
Thus, our method can be easily generalized to describe the structural
deformations of a 2D planar structure, not only under uniaxial strain, but
under any strain condition.

In order to minimize $\Lambda$ in Eq.~(\ref{Lagrange}), with respect to
the bond stretching and angle bending
deformations, one needs to express $\delta L$ in terms of these deformations.

\subsection{$\delta L$ as a function of bond deformations}
Without loss of generality, we may assume that the structure is periodic.
A non-periodic (i.e. amorphous) structure could be considered as periodic
with infinite periodicity. For convenience, let us assume that the unit cell
vectors for $\varepsilon=0$ are $\mathbf{a}_0=a_0\hat{\mathbf{i}}$ and 
$\mathbf{b}_0=b_{x0}\hat{\mathbf{i}}+b_{y0}\hat{\mathbf{j}}$, as shown in
Fig.~\ref{projections}. Let us apply tensile strain by stretching the
structure along the line connecting two equivalent atoms in different unit
cells. The vector connecting those two atoms, (which determine the strain
direction), is $\mathbf{L}_0=n\mathbf{a}_0+m\mathbf{b}_0$, where $n$ and
$m$ are integers. Under the applied strain the vector $\mathbf{L}_0$ will
be deformed to $\mathbf{L}$, so that the vectors $\mathbf{L}$ and
$\mathbf{L}_0$ are parallel, i.e. $\mathbf{L}_0$ will be just
elongated. The unit cell vectors $\mathbf{a}_0$ and $\mathbf{b}_0$ will
be also deformed to $\mathbf{a}$ and $\mathbf{b}$, respectively, so that
$\mathbf{L}_0 = n\mathbf{a}_0+m\mathbf{b}_0 \parallel 
\mathbf{L} = n\mathbf{a}+m\mathbf{b}$.

If $L_0$ and $L=L_0+\delta L$ are the lengths of the vectors $\mathbf{L}_0$
and $\mathbf{L}$, respectively, and $\hat{\boldsymbol{\varepsilon}}$ is the unit
vector directed along the strain direction (i.e. $\hat{\boldsymbol{\varepsilon}}=
(n\mathbf{a}_0+m\mathbf{b}_0)/(n^2a_0^2+m^2b_0^2+2nm\mathbf{a}_0\mathbf{b}_0)^{1/2}$,
where $b_0=(b_{x0}^2+b_{y0}^2)^{1/2}$), then 
$L=\hat{\boldsymbol{\varepsilon}}(n\mathbf{a}+m\mathbf{b})=
n(\hat{\boldsymbol{\varepsilon}}\mathbf{a})+m(\hat{\boldsymbol{\varepsilon}}\mathbf{b})$
and $L_0=\hat{\boldsymbol{\varepsilon}}(n\mathbf{a_0}+m\mathbf{b_0})=
n(\hat{\boldsymbol{\varepsilon}}\mathbf{a_0})+m(\hat{\boldsymbol{\varepsilon}}\mathbf{b_0})$,
i.e. $L$ ($L_0$) depend on the projections of $\mathbf{a}$, and $\mathbf{b}$
($\mathbf{a}_0$ and $\mathbf{b}_0$) on the strain direction. 

The vectors $\mathbf{a}_0$ and $\mathbf{b}_0$ can be expressed as a sum
of bond vectors $\mathbf{r}_{0ai}$ and $\mathbf{r}_{0bi}$, respectively,
($i=1,2,3,...$), which correspond to specific bonds of the undeformed
structure, constituting a crooked line connecting the tails of $\mathbf{a}_0$
and $\mathbf{b}_0$ with their heads, i.e. $\mathbf{a}_0=\sum_i\mathbf{r}_{0ai}$
and $\mathbf{b}_0=\sum_i\mathbf{r}_{0bi}$. Thus, if the bond vectors 
$\mathbf{r}_{0ai}$ and $\mathbf{r}_{0bi}$ are deformed under strain into
$\mathbf{r}_{ai}$ and $\mathbf{r}_{bi}$, respectively, then 
$\mathbf{a}=\sum_i\mathbf{r}_{ai}$ and $\mathbf{b}=\sum_i\mathbf{r}_{bi}$.
This is shown schematically in Fig.~\ref{projections}, where the sum
of the red colored vectors, (denoted as $\mathbf{r}_{ai}$, $i=1,2,3,...$),
constitute $\mathbf{a}$, while the sum of the green colored vectors, (denoted
as $\mathbf{r}_{bi}$, $i=1,2,3,...$), constitute $\mathbf{b}$. Obviously,
the corresponding sums of the projections of $\mathbf{r}_{ai}$ and
$\mathbf{r}_{bi}$ along the strain direction equals the projection of
$\mathbf{a}$ and $\mathbf{b}$, respectively, along the same direction.
These projections of $\mathbf{r}_{ai}$ and $\mathbf{r}_{bi}$ are shown as
black arrows in Fig.~\ref{projections}, and should be considered as
positive or negative. Thus,
\begin{eqnarray}\label{sum}
\delta L & = & L-L_0 \\ \nonumber
         & = & n\sum_i\left(\hat{\boldsymbol{\varepsilon}}\mathbf{r}_{ai}-\hat{\boldsymbol{\varepsilon}}\mathbf{r}_{0ai}\right)+
               m\sum_i\left(\hat{\boldsymbol{\varepsilon}}\mathbf{r}_{bi}-\hat{\boldsymbol{\varepsilon}}\mathbf{r}_{0bi}\right),
\end{eqnarray}
i.e. $\delta L$ can be expressed as a function of the differences of the
projections of the $\mathbf{r}_{0ai}$, $\mathbf{r}_{ai}$ and the
$\mathbf{r}_{0bi}$, $\mathbf{r}_{bi}$ vectors, along the strain direction.
We should note that, although the vectors, $\mathbf{a}$, $\mathbf{b}$,
$\mathbf{a}_0$, $\mathbf{b}_0$ are not uniquely expressed in terms of 
bond vectors, the sums of the projections are unique and one could always 
choose optimal paths (e.g. of minimal length) of bond vectors.
Let us now see how the differences of those projections depend on the bond
deformations.

\subsection{The strain constrain} 

Let us assume that strain along a specific direction is applied to a bond,
as shown in Fig.~\ref{Fig1}(b). For convenience we have assumed that
the strain direction coincides with the x-axis direction. Let us further
assume that at equilibrium for $\varepsilon=0$, the bond length and the
angle between the bond and the strain direction are $l_0$ and $\theta_0$,
and under strain they become $\theta_0+\delta\theta$ and $l_0+\delta l$, 
respectively. If the projections of the bond along and normal to the strain
direction for $\varepsilon=0$ are $x_0$ and $y_0$, respectively, and under
strain they are $x_0+\delta x$ and $y_0+\delta y$, respectively, then
$x_0=l_0\cos\theta_0$, $y_0=l_0\sin\theta_0$,
$x_0+\delta x=(l_0+\delta l)\cos(\theta_0+\delta\theta)$ and
$y_0+\delta y=(l_0+\delta l)\sin(\theta_0+\delta\theta)$.

Thus the projection of the bond deformation along the strain direction is
\begin{eqnarray}\label{x_deform}
\delta x 
         & \approx & \delta l\cos\theta_0-l_0\sin\theta_0\delta\theta
\end{eqnarray}
and the projection normal to the strain direction is
\begin{eqnarray}\label{y_deform}
\delta y 
         & \approx & \delta l\sin\theta_0+l_0\cos\theta_0\delta\theta.
\end{eqnarray}

According to Eq.~(\ref{x_deform}), the projection $\delta x$ of the deformation
of $\mathbf{r}_{0ai}$ along the strain direction $\hat{\boldsymbol{\varepsilon}}$ is
\begin{eqnarray}
\delta x & = & \hat{\boldsymbol{\varepsilon}}\mathbf{r}_{ai}-\hat{\boldsymbol{\varepsilon}}\mathbf{r}_{0ai} \\ \nonumber
         & = & \delta l_{ai}\cos\theta_{0ai}-l_{0ai}\sin\theta_{0ai}\delta\theta_{ai},
\end{eqnarray}
where $l_{0ai}=|\mathbf{r}_{0ai}|$, $\theta_{0ai}$ is the angle between
$\mathbf{r}_{a0i}$ and the strain direction (i.e. 
$\cos\theta_{0ai}=\hat{\boldsymbol{\varepsilon}}\mathbf{r}_{0ai}/l_{0ai}$), and
$\delta l_{ai}$ and $\delta\theta_{ai}$ are the deformations of $l_{0ai}$
and $\theta_{0ai}$, respectively. Changing the index "$a$" with "$b$", we get
the corresponding relation for $\mathbf{r}_{0bi}$. Consequently,
\begin{eqnarray}
\delta L  & = & n\sum_i\left(\delta l_{ai}\cos\theta_{0ai}-l_{0ai}\sin\theta_{0ai}\delta\theta_{ai}\right)+ \\ \nonumber
& & m\sum_i\left(\delta l_{bi}\cos\theta_{0bi}-l_{0bi}\sin\theta_{0bi}\delta\theta_{bi}\right).
\end{eqnarray}
As a function of the projections of independently deformed bonds, this
equation is written as
\begin{equation}
\delta L = \sum_i q_i\left(\delta l_{i}\cos\theta_{0i}-l_{0i}\sin\theta_{0i}\delta\theta_{i}\right)
\end{equation}
where here index $i$ is the same as in Eq.~(\ref{energy2}), (i.e. it
runs over the bond vectors of the independently deformed bonds) and
$q_i$ is the number of the bond vectors $\mathbf{r}_{0a}$ and $\mathbf{r}_{0b}$
with equivalent deformations, which contribute to
the sums in Eq.~(\ref{sum}). Obviously, if $\mathbf{r}_{i}$ does not
contribute to the sums in Eq.~(\ref{sum}), then $q_i=0$, and if
$-\mathbf{r}_{i}$ contributes to the sums in (\ref{sum}) instead of
$\mathbf{r}_{i}$, then the angle $\theta_{0i}$ of the above equation should
be replaced by $\theta_{0i} +\pi$, which changes the sign of both
$\cos\theta_{0i}$ and $\sin\theta_{0i}$. This sign change can be absorbed in
$q_i$, and therefore, the constrain of our case has the form 
\begin{equation}\label{constrain}
\varepsilon - \sum_i q_i\left(\delta l_{i}\cos\theta_{0i}-l_{0i}\sin\theta_{0i}\delta\theta_{i}\right)/L_0 = 0.
\end{equation}

As one can see, the deformation energy in Eq.~(\ref{energy2}) is expressed
as a function of the deformations $\delta l_i$ and $\delta\phi_{ij}$, while
the constrain in Eq.~(\ref{constrain}) is expressed as a function of
$\delta l_i$ and $\delta\theta_i$. As we show in the 
Sec.~\ref{delta_phi_delta_theta}, 
\begin{equation}\label{square_delta}
\forall \phi_{ij} \in (0,\pi], \quad \delta \phi_{ij}^2=(\delta\theta_j-\delta\theta_i)^2,
\end{equation}
and therefore, the function $\Lambda$ in Eq.~(\ref{Lagrange}), which
has to be minimized, can be rewritten as
\begin{eqnarray}\label{Lambda}
\Lambda & = & \Lambda\left(\{\delta l_i\},\{\delta\theta_{i}\},\lambda\right) \\
        & = & \frac{1}{2}\sum_i \left(n_i k_{s,i}\delta l_i^2 + \frac{1}{2}\sum_j m_{ij} k_{b,ij}(\delta\theta_{i}-\delta\theta_{j})^2\right)  \nonumber \\
  &   & + \lambda\left(\varepsilon-\sum_i q_i\left(\delta l_{i}\cos\theta_{0i}-l_{0i}\sin\theta_{0i}\delta\theta_{i}\right)/L_0\right), \nonumber
\end{eqnarray}
where by $\{\delta l_i\}$ and $\{\delta \theta_{i}\}$ we denote all the
$\delta l_i$ and $\delta\theta_i$ independent variables, respectively,
(i.e. $\{\delta l_i\}=\delta l_1, \delta l_2,\ldots$ and 
$\{\delta \theta_{i}\}=\delta\theta_1, \delta\theta_2,\ldots$), and
therefore $\Lambda$ becomes a function of only $\delta l_i$,
$\delta\theta_i$ and $\lambda$.

It is worth noting that the projection of 
$\delta \mathbf{L}=\mathbf{L}-\mathbf{L}_0$ normal to the strain direction
should be zero, i.e. (according to Eq.~(\ref{y_deform}))
\begin{equation}\label{normal_proj}
\sum_i q_i(\delta l_i\sin\theta_{0i}+l_{0i}\cos\theta_{0i}\delta\theta_i)=0.
\end{equation}
As we will see, minimizing $\Lambda$ in (\ref{Lambda}) we will be able to calculate 
the differences of $\delta\theta_i$ for the same atom, (i.e. the bond angle
deformations $\delta\phi_{ij}$), but not the deformations $\delta\theta_i$
themselves, which give the direction of the bonds with respect to the strain
direction. However, using (\ref{normal_proj}) and the results of the
minimization in (\ref{Lambda}), the deformations $\delta\theta_i$ can
be also determined and we can have a complete figure for the deformations
of the structure. 

\section{Minimization of $\Lambda(\{\delta l_i\},\{\delta \theta_{ij}\},\lambda)$ }

The steady state of $\Lambda$ occurs at the specific $\delta l_i$ and
$\delta\theta_i$ values for which 
\begin{equation}\label{partial}
\partial \Lambda/\partial \delta l_i=0 \qquad \textrm{and} \qquad  
\partial \Lambda/\partial \delta \theta_i=0.
\end{equation}
$\delta l_i$ appears only in one term of $U$, namely in
$(1/2)k_{s,i}\delta l_i^2$. Consequently, from 
$\partial \Lambda/\partial \delta l_i=0$ we obtain
\begin{equation}\label{length}
\delta l_i = \frac{\lambda}{L_0} \frac{q_i}{n_i}\frac{\cos\theta_{0i}}{k_{s,i}}.
\end{equation}
On the other hand, $\delta \theta_i$ appears in 4 terms of $U$ 
(see Fig.~\ref{Fig1}(a)), namely in
$m_{ij_1}k_{b,ij_1}(\delta\theta_i-\delta\theta_{j_1})^2$ and 
$m_{ij_2}k_{b,ij_2}(\delta\theta_i-\delta\theta_{j_2})^2$ for the angles 
$\delta \phi_{ij_1}$ and $\delta \phi_{ij_2}$ of atom A, and
$m_{ij_3}k_{b,ij_3}(\delta\theta_i-\delta\theta_{j_3})^2$ and 
$m_{ij_4}k_{b,ij_4}(\delta\theta_i-\delta\theta_{j_4})^2$
for the angles $\delta \phi_{ij_3}$ and $\delta \phi_{ij_4}$ of atom B.
From $\partial \Lambda/\partial \delta\theta_i=0$ we obtain the linear system
\begin{equation}\label{theta}
\frac{1}{2}\sum_{k=1}^4 m_{ij_k}k_{b,ij_k}(\delta\theta_i-\delta\theta_{j_k})=-\lambda q_i l_{0i} \sin\theta_{0i}/L_0.
\end{equation}

Substituting the expressions for $\delta\theta_i$ obtained from
Eq.~(\ref{theta}) and  the expressions for $\delta l_i$ shown in
Eq.~(\ref{length}) into (\ref{constrain}), we obtain an equation
for $\lambda$. Solving this equation with respect to $\lambda$, we
obtain $\lambda$ as a function of the strain $\varepsilon$ and the
strain angle $\theta_0$.

As we show in the Section~\ref{Physical_meaning},
\begin{equation}\label{U_min}
U_{min}=\lambda\varepsilon/2,
\end{equation} 
where $U_{min}$ is the minimum of $U$ subject to the constrain
$\varepsilon=\delta L/L_0$. Thus, if $\lambda$ is determined, then
$U_{min}$ can also be determined. Eq.~(\ref{U_min}) gives a physical
meaning in the Lagrange multiplier $\lambda$ and minimizes the effort
to find a convenient expression for $U_{min}$ as a function of
$k_{s,i}$ and $k_{b,ij}$ for strain $\varepsilon$.

\section{Including second nearest neighbor stretching terms }
As we can see from Eqs.~(\ref{length}) and (\ref{theta}), the original
stick and spiral model, expressed utilizing (\ref{energy2}), does
not provide any coupling between $\delta l_i$ and $\delta\phi_{ij}$.
However, as already mentioned, including energy terms which describe
stretching from second nearest neighbor interactions, we obtain a more
accurate model, since it provides coupling between $\delta l_i$ and
$\delta \phi_{ij}$.

Let us assume that atoms B and C are second nearest neighbors, forming
bonds $i$ and $j$, respectively, with atom A. If $\mathbf{r}_{0i}$ and
$\mathbf{r}_{0j}$ are the bond vectors of bonds $i$ and $j$, at equilibrium
for $\varepsilon=0$, then, depending on the orientation of $\mathbf{r}_{0i}$
and $\mathbf{r}_{0j}$, the interatomic distance $r_{0ij}$ between atoms
B and C is either the magnitude of the vector 
$\mathbf{r}_{0j}-\mathbf{r}_{0i}$ (if both heads or tails of
$\mathbf{r}_{0i}$ and $\mathbf{r}_{0j}$ are at the position of atom A),
or the vector $\mathbf{r}_{0j}+\mathbf{r}_{0i}$ (if the tail of the one
and the head of the other are at the position of atom A). 

If the interatomic distance $r_{0ij}$ is deformed upon strain
by $\delta r_{ij}$, then the deformation energy per unit cell $U$ is
\begin{equation}\label{energy3}
U=U_1+U_2=U_1+(1/2)\sum_i\sum_j p_{ij}(1/2)k_{s,ij}\delta r_{ij}^2,
\end{equation} 
where  $U_1$ is the deformation energy
of the original stick and spiral model in Eq.~(\ref{energy2}) and $U_2$
describes the contribution due to stretching deformations of second nearest 
neighbor interatomic distances.
The factor $1/2$ in the second term of Eq.~(\ref{energy3}) is inserted
to avoid double counting, the notation $i$ and $j$ is the same as in
(\ref{energy2}) and $p_{ij}$ is the number of the equivalent second
nearest neighbor interatomic distances in the unit cell with a
$\delta r_{ij}$ deformation. Obviously, $p_{ij}=m_{ij}$, because each
specific bond angle $\phi_{ij}$ corresponds to a specific second nearest
neighbor interatomic distance $r_{ij}$. 

Consequently, for the atomic arrangement shown in Fig.~\ref{Fig1}(a),
Eqs.~(\ref{length}) and (\ref{theta}) should be replaced by
\begin{equation}\label{length2}
n_ik_{s,i}\delta l_i+\frac{1}{2}\sum_{k=1}^4 m_{ij_k}k_{s,ij_k}\delta r_{ij_k}\frac{\partial\delta r_{ij_k}}{\partial\delta l_i}=\lambda q_i\cos\theta_{0i}/L_0
\end{equation}
and
\begin{eqnarray}\label{theta2}
& & \frac{1}{2}\sum_{k=1}^4 m_{ij_k}\left[k_{b,ij_k}(\delta\theta_i-\delta\theta_{j_k}) 
    + k_{s,ij_k}\delta r_{ij_k}\frac{\partial\delta r_{ij_k}}{\partial\delta \theta_i}\right] \nonumber \\
& & =-\lambda q_i l_{0i} \sin\theta_{0i}/L_0,
\end{eqnarray}
which have to be solved.

As we show in the Sec.~\ref{delta_r_ij}, 
\begin{eqnarray}\label{second_fin}
r_{0ij}\delta r_{ij} & = & \left(l_{0i} \mp l_{0j}\cos(\theta_{0i}-\theta_{0j})\right)\delta l_i \nonumber \\
                     &   & +\left(l_{0j} \mp l_{0i}\cos(\theta_{0j}-\theta_{0i})\right)\delta l_j \nonumber \\
                     &   & \pm l_{0i} l_{0j}\sin(\theta_{0i}-\theta_{0j})(\delta\theta_{i}-\delta\theta_{j}),
\end{eqnarray}
and
\begin{equation}\label{partial_l_fin}
\partial\delta r_{ij}/\partial\delta l_i = \left[l_{0i} \mp l_{0j}\cos(\theta_{0i}-\theta_{0j})\right]/r_{0ij},
\end{equation}
\begin{equation}\label{partial_theta_fin}
\partial\delta r_{ij}/\partial\delta\theta_i = \pm l_{0i} l_{0j}\sin(\theta_{0i}-\theta_{0j})/r_{0ij}.
\end{equation}
The upper signs, (wherever $\pm$ and $\mp$ appear), occur when
$\mathbf{r}_i$ and $\mathbf{r}_j$ have their tails (or their heads)
at the position of the same atom and the lower signs, when the tail
of the one and the head of the other are at the position of the same
atom, as explained in Sec.~\ref{delta_r_ij}.

Obviously, if $k_{s,ij}=0$, then $U_2=0$ and the modified stick and
spiral model reduces to the original one. Thus, we can treat both
models by solving the system of Eqs.~(\ref{length2}) and (\ref{theta2})
of the modified model. Then, by setting $k_{s,ij}=0$ in these solutions,
we directly get the solutions of (\ref{length}) and (\ref{theta})
of the original model. This is the subject of the next section specified for graphene.

\section{Application to graphene}\label{app_graphene}
\begin{figure}[!tb]
\begin{center}
\includegraphics[width=0.475\textwidth, clip]{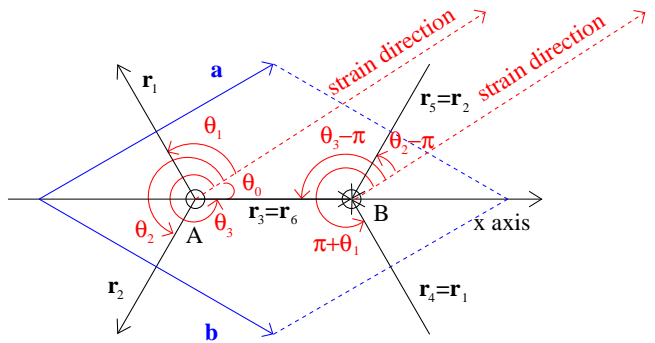}
\end{center}
\caption{(Color online) 
Graphene unit cell. The lattice vectors are $\mathbf{a}=\mathbf{r}_1-\mathbf{r}_3$
and $\mathbf{b}=\mathbf{r}_1-\mathbf{r}_2$. The bond vectors for atom A are $\mathbf{r}_1$,
$\mathbf{r}_2$ and $\mathbf{r}_3$, while for atom B they are $-\mathbf{r}_1$, $-\mathbf{r}_2$
and $-\mathbf{r}_3$. The bond angles $\theta_i$ with respect to the strain direction are 
also shown.}
\label{graphene_unit}
\end{figure}

Bellow, as well as in the appendices, whenever the indices $i^\prime$,
$j^\prime$ and $k^\prime$ are used, $(i^\prime, j^\prime, k^\prime)=(1,2,3)$,
or $(2,3,1)$, or $(3,1,2)$.

\subsection{The energy}
Fig.~\ref{graphene_unit} shows the unit cell of graphene, which is
defined by the lattice vectors 
$\mathbf{a}=(\sqrt{3}/2)(\sqrt{3}\hat{\mathbf{i}}+\hat{\mathbf{j}})a_0$
and $\mathbf{b}=(\sqrt{3}/2)(\sqrt{3}\hat{\mathbf{i}}-\hat{\mathbf{j}})a_0$,
where $a_0$ is the bond length of graphene. In this figure, A and B are the
2 atoms of the lattice base. As one can see, there are 3 bonds per unit cell,
which can be deformed independently, corresponding to the bond vectors
$\mathbf{r}_1$, $\mathbf{r}_2$ and $\mathbf{r}_3$ of atom A, or the bond
vectors $\mathbf{r}_4=\mathbf{r}_1$, $\mathbf{r}_5=\mathbf{r}_2$ and
$\mathbf{r}_6=\mathbf{r}_3$ of atom B. Consequently, in
Eqs.~(\ref{energy2}) and (\ref{energy3}), $n_i=1$ ($i=1,2,3$).
Moreover, as one can see in Fig.~\ref{graphene_unit}, there are six
bond angles (with respect to the strain direction) $\theta_i$ per unit
cell. Three of them correspond to atom A and three to atom B. Since the
bond vectors of atom A and B are the same, the angles $\theta_i$
corresponding to the bonds of atom A are the same with those corresponding
to atom B. Consequently, only three of those six angles can be considered
as independently deformed, and $m_{ij}=2$. 
Moreover, due to symmetry reasons, $k_{s,i}=k_{s1}$, $k_{s,ij}=k_{s2}$
and $k_{b,ij}=k_b$.

Thus, the energy per unit cell in the original stick and spiral model
(according to Eq.~(\ref{energy2})) is
\begin{eqnarray}\label{grgr1}
U = U_1 & = & \frac{1}{2}k_{s1}\left(\delta l_1^2+\delta l_2^2+\delta l_3^2\right) 
        + k_b^\prime a_0^2\left[(\delta\theta_1-\delta\theta_2)^2+ \right. \nonumber \\
  &   & \left.(\delta\theta_2-\delta\theta_3)^2+(\delta\theta_3-\delta\theta_1)^2\right],
\end{eqnarray}
where $k_b^\prime=k_b/a_0^2$.

In the unit cell of graphene shown in Fig.~\ref{graphene_unit}, there
are six second nearest neighbor interatomic distances, namely $r_{12}$,
$r_{23}$, $r_{31}$, $r_{45}$, $r_{56}$ and $r_{64}$, where
$r_{45}=r_{12}$, $r_{56}=r_{23}$ and $r_{64}=r_{31}$. Consequently,
there are only three second nearest neighbor interatomic distances,
which can be deformed independently and $U_2$ in Eq.~(\ref{energy3}) is
\begin{equation}\label{grgr2}
U_2=k_{s2}(\delta r_{1,2}^2+\delta r_{2,3}^2+\delta r_{3,1}^2),
\end{equation} 
where $\delta r_{ij}$ are given by (\ref{second_fin}), and therefore,
the energy per atom $U$ in the modified model is $U=U_1+U_2$.

\subsection{The strain constrain}\label{strain_constr}
As a function of the independently deformed bond vectors $\mathbf{r}_i$,
the unit cell vectors $\mathbf{a}$ and $\mathbf{b}$ can be written as
\begin{equation}\label{a_b}
\mathbf{a}=\mathbf{r}_3-\mathbf{r}_2 \qquad \textrm{and} \qquad 
\mathbf{b}=\mathbf{r}_3-\mathbf{r}_1.
\end{equation}
Thus, if $\mathbf{L}_0=n\mathbf{a}+m\mathbf{b}$ defines the strain
direction, then $\mathbf{L}_0=(n+m)\mathbf{r}_3-n\mathbf{r}_2-m\mathbf{r}_1$,
and consequently the $q_i$s in (\ref{constrain}) are $q_3=n+m$, $q_2=-n$
and $q_1=-m$. As we show in the Sec.~\ref{graphene_strain},
\begin{equation}\label{q_s}
q_i=2L_0/(3a_0)\cos\theta_{0i},
\end{equation}
where
\begin{equation}\label{theta_0i}
\theta_{0i}=2\pi i/3-\theta_0, \quad i=1,2,3,
\end{equation}
and consequently, (as shown in the same Appendix), the strain constraint of Eq.~(\ref{constrain}) 
takes the form
\begin{equation}\label{gr_strain2}
\varepsilon = \frac{2}{3a_0}\sum_{j=1}^3\cos^2\theta_{0j}\delta l_j- 
         \frac{1}{3}\sum_{j=1}^3\sin2\theta_{0j}(\delta\theta_j-\delta\theta_i),
\end{equation}
while (\ref{normal_proj}) becomes
\begin{equation}\label{normal_proj_gr}
\delta\theta_i=\sum_{j=1}^3\left(\frac{2}{3}\cos^2\theta_{0j}(\delta\theta_i-\delta\theta_j)-\frac{\delta l_j}{3a_0}\sin 2\theta_{0j}\right),
\end{equation}
respectively, where $i=1$, or 2, or 3.

\subsection{Solving for the deformations $\delta l_i$ and $\delta\theta_i$}
As we show in the Sec.~(\ref{modified}), Eqs.~(\ref{length2})
and (\ref{theta2}) give
\begin{eqnarray}\label{gr_2_length}
& & (\sqrt{3}/2)k_{s2}\left[\sqrt{3}(2\delta l_{i^\prime}+\delta l_{j^\prime}+\delta l_{k^\prime})+a_0(\delta\theta_{j^\prime}-\delta\theta_{k^\prime})\right] \nonumber \\
& + & k_{s1}\delta l_{i^\prime}=(2\lambda/3a_0)\cos^2\theta_{0i^\prime}
\end{eqnarray}
and
\begin{eqnarray}\label{gr_2_angle}
& & \left(k_b^\prime+k_{s2}/4\right)a_0^2[(\delta\theta_{i^\prime}-\delta\theta_{j^\prime})+(\delta\theta_{i^\prime}-\delta\theta_{k^\prime})] \nonumber \\
& + & (\sqrt{3}/4)a_0k_{s2}(\delta l_{k^\prime}-\delta l_{j^\prime})=-(\lambda/6)\sin2\theta_{0i^\prime}.
\end{eqnarray}
The solution of these equations, (as shown in the same appendix),
is of the form 
\begin{equation}\label{delta_l_i}
\delta l_i=3a_0(\xi_1^\prime\cos^2\theta_{0i}+\xi_2^\prime)
\end{equation}
and 
\begin{equation}\label{delta_theta_ij}
\delta \theta_j - \delta\theta_i = \xi_3^\prime(\sin2\theta_{0i}-\sin2\theta_{0j}),
\end{equation}
where $\xi_1^\prime=8k_b^\prime\lambda/(9a_0^2K^\prime)$, 
$\xi_2^\prime=k_{s2}\lambda(k_{s1}-18k_b^\prime)/[9a_0^2K^\prime (k_{s1}+6k_{s2})]$, 
$\xi_3^\prime=2\lambda k_{s1}/(9a_0^2K^\prime)$ and 
$K^\prime=k_{s1}k_{s2}+(4k_{s1}+6k_{s2})k_b^\prime$.
For these expressions of $\delta l_i$ and
$\delta \theta_j - \delta\theta_i$, Eqs.~(\ref{gr_strain2}) and 
(\ref{normal_proj_gr}) yield
\begin{equation}\label{xi_dependence}
\varepsilon=(9\xi_1^\prime+12\xi_2^\prime+2\xi_3^\prime)/4 \quad \textrm{and} \quad
\delta\theta_i=-\xi_3^\prime\sin 2\theta_{0i},
\end{equation}
(see Sec.~\ref{modified} for details). Consequently,
\begin{equation}
\varepsilon = \lambda K_0/[9a_0^2K^\prime(k_{s1}+6k_{s2})], 
\end{equation}
where $K_0=k_{s1}^2+9k_{s1}k_{s2}+18(k_{s1}+3k_{s2})k_b^\prime$
and therefore,
\begin{equation}\label{lambda_mod_mod}
\lambda = 9a_0^2\varepsilon K^\prime (k_{s1}+6k_{s2})/K_0.
\end{equation}

Thus, 
\begin{equation}\label{l_lambda}
\delta l_i=3a_0\lambda_i \varepsilon, \quad \textrm{and} \quad 
\delta\theta_j-\delta\theta_i=\mu_{ij}\varepsilon,
\end{equation}
where
\begin{equation}\label{lambda_i}
\lambda_i=\xi_1\cos^2\theta_{0i}+\xi_2,
\end{equation}
\begin{equation}\label{mu_ij}
\mu_{ij}=-\mu_{ji}=\xi_3(\sin2\theta_{0i}-\sin2\theta_{0j})
\end{equation}
and
\begin{equation}\label{xi_1}
\xi_1=8k_b^\prime(k_{s1}+6k_{s2})/K_0,
\end{equation}
\begin{equation}\label{xi_2}
\xi_2=k_{s2}(k_{s1}-18k_b^\prime)/K_0,
\end{equation}
\begin{equation}\label{xi_3}
\xi_3=2k_{s1}(k_{s1}+6k_{s2})/K_0.
\end{equation}
Using Eq.~(\ref{cos_sin_sin_2}), (\ref{mu_ij}) gives
\begin{equation}\label{mu_ij_second}
\mu_{i^\prime j^\prime}=-\mu_{j^\prime i^\prime}=-\sqrt{3}\xi_3\cos2\theta_{0k^\prime}
\end{equation}

Obviously, Eq.~(\ref{xi_dependence}) leads to
\begin{equation}\label{xi_dependence_2}
9\xi_1+12\xi_2+2\xi_3=4 \quad \textrm{and} \quad \delta\theta_i=-(\xi_3\sin 2\theta_{0i})\varepsilon.
\end{equation}
The former shows that $\xi_1$, $\xi_2$ and $\xi_3$ are not independent.

Moreover, according to the relations between $\phi_{ij}$ and $\theta_i$ 
shown in Sec.~\ref{delta_phi_delta_theta}, the relations between the
$\phi_{ij}$ and $\theta_i$ angles of graphene, shown in Fig.~\ref{graphene_unit} are
\begin{equation}
\phi_{21}=\theta_2-\theta_1,  \quad \phi_{32}=\theta_3-\theta_2 \quad \textrm{and} \quad \phi_{13}=2\pi+\theta_1-\theta_3.
\end{equation}
Thus, the bond angle deformations $\delta\phi_{ij}$ are 
\begin{equation}
\delta\phi_{i^\prime j^\prime}=\delta\phi_{j^\prime i^\prime}=\delta\theta_{j^\prime}-\delta\theta_{i^\prime}.
\end{equation}

Due to the symmetry of the unit cell, the results we find for strain angle
$\theta_0$, will be the same for strain angles $n\pi/3\pm\theta_0$, 
$n=0,1,2,3,4,5$. Thus, without loss of generality, we may assume that 
$0\le\theta_0\le\pi/6$.

\begin{figure*}[!tb]
\begin{center}
\includegraphics[width=0.95\textwidth,clip]{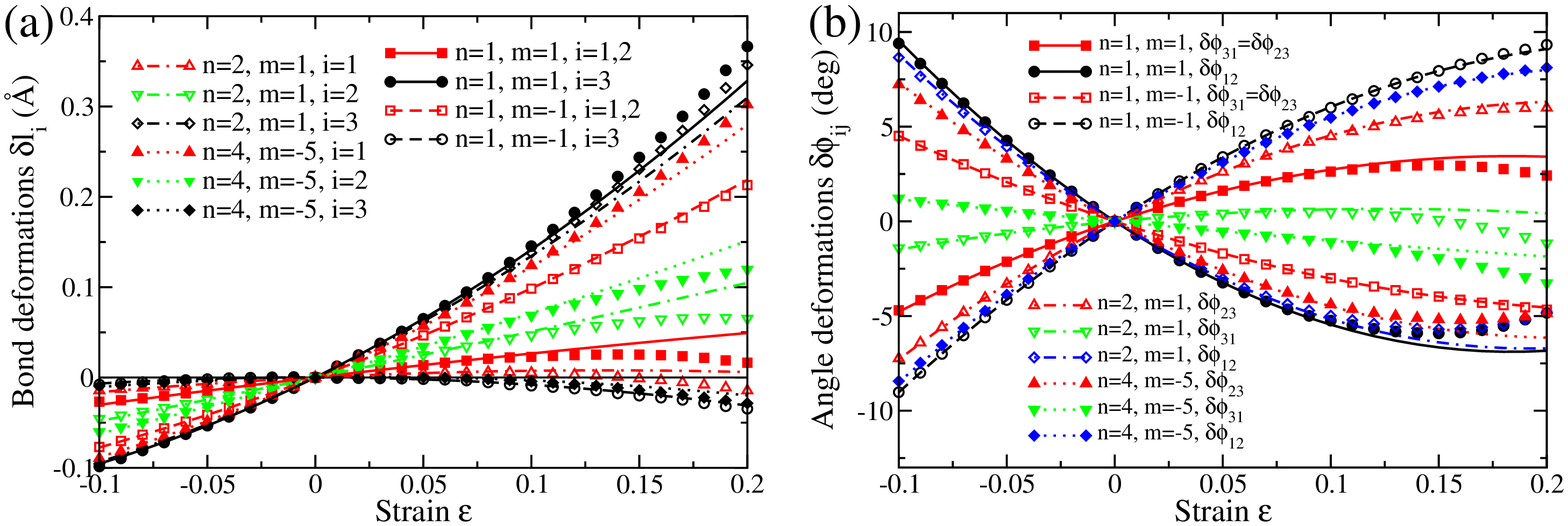}
\end{center}
\caption{(Color online)
(a) Bond length deformations $\delta l_i$ and (b) bond angle
deformations $\delta \phi_{ij}$ as a function of strain $\varepsilon$, 
upon stretching along the directions
defined by the vectors $\mathbf{L}=n\mathbf{a}+m\mathbf{b}$.
$n=1$ and $m=1$ corresponds to the arm chair direction.
$n=1$ and $m=-1$ corresponds to the zig-zag direction.}
\label{deformations}
\end{figure*}

\subsection{Energy, Young's modulus and Poisson's ratio}

\begin{figure}
\begin{center}
\includegraphics [width=0.475\textwidth,clip]{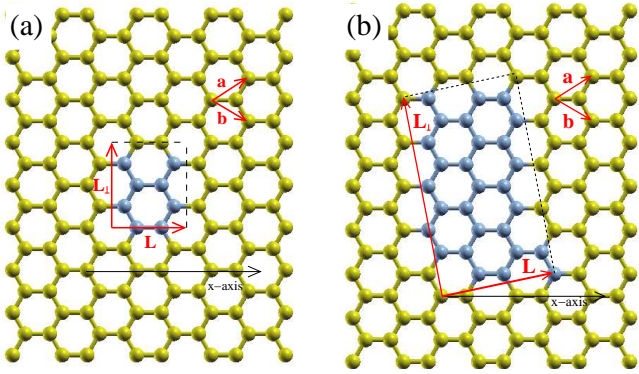}
\end{center}
\caption{\label{Fig4} (Color online)
Rectangular unit cells and strain directions used in our calculations.
Unit cell atoms are shown with blue color.
(a) For strain along the arm chair ($\mathbf{L}=\mathbf{a}+\mathbf{b}$)
and the zig-zag ($\mathbf{L_\perp}=\mathbf{a}-\mathbf{b}$) direction, and
(b) for strain along the direction of the vectors
$\mathbf{L}=2\mathbf{a}+\mathbf{b}$ and 
$\mathbf{L_\perp}=4\mathbf{a}-5\mathbf{b}$.
}
\end{figure}

According to Eq.~(\ref{U_min}), the deformation energy per unit cell
is $U=\lambda\varepsilon/2$. For graphene, $\lambda$ is given by 
(\ref{lambda_mod_mod}), and consequently, 
\begin{equation}\label{gr_energy}
U=(3a_0\varepsilon)^2A, 
\end{equation}
where
\begin{equation}\label{energy_gr2}
A=(k_{s1}+6k_{s2})(k_{s1}k_{s2}+(4k_{s1}+6k_{s2})k_b^\prime)/(2K_0).
\end{equation}

As for the Young's modulus $E$, it is easy to show that
$E=2U/(V\varepsilon^2)$, 
where $V$ is the volume of the unit cell ($V=3\sqrt{3}a_0^2d_0/2$) and $d_0$
is the hypothetical depth of the graphene layer, which is assumed to be equal
to the graphite interlayer separation ($d_0=3.34$\AA), in order to direct
compare the Young's modulus values of two dimensional (2D) carbon structures
with the known values for three dimensional (3D) systems, like 
graphite \cite{PCCP_Fthenakis}. 
Thus, for the above expression for $A$,
\begin{equation}\label{gr_Young}
E=4\sqrt{3}A/d_0,
\end{equation}

Moreover, in Sec.~\ref{Poisson} we show that the Poisson's ratio
$\nu$ is
\begin{equation}\label{nu}
\nu = -3\xi_1/4-3\xi_2+\xi_3/2,
\end{equation}
which for the $\xi_1$, $\xi_2$ and $\xi_3$ expressions of 
(\ref{xi_1}), (\ref{xi_2}) and (\ref{xi_3}) becomes
\begin{equation}\label{nu_k}
\nu = [(k_{s1}+6k_{s2})(k_{s1}-6k_b^\prime)-3k_{s2}(k_{s1}-18k_b^\prime)]/K_0.
\end{equation}
As one can see from the above expressions,
$U$, $E$ and $\nu$ are independent of the strain angle $\theta_0$, and consequently,
graphene is isotropic.

\subsection{Relations between $k_{s1}$, $k_{s2}$ and $k_b^\prime$
with $\xi_1$, $\xi_2$, $\xi_3$ and $A$}
One would have thought that Eqs.~(\ref{xi_1}), (\ref{xi_2}) and (\ref{xi_3}), which
form a $3\times 3$ system of equations, would provide solutions
for $k_{s1}$, $k_{s2}$ and $k_b^\prime$ as functions of
$\xi_1$, $\xi_2$ and $\xi_3$. 
However, as shown in Eq.~(\ref{xi_dependence_2}), $\xi_1$, $\xi_2$ and $\xi_3$
are not independent, and therefore, these equations can not provide relations for
$k_{s1}$, $k_{s2}$ and $k_b^\prime$ as functions of $\xi_1$, $\xi_2$
and $\xi_3$. On the other hand, $A$, which is independent
of $\xi_1$, $\xi_2$ and $\xi_3$, is also a function of 
$k_{s1}$, $k_{s2}$ and $k_b^\prime$. Therefore, $k_{s1}$, $k_{s2}$ and $k_b^\prime$
could be written as functions of $\xi_1$, $\xi_2$, $\xi_3$ and $A$.

As we show in the Sec.~\ref{relations},
\begin{equation}\label{relations1}
\frac{k_b^\prime}{k_{s1}}=\frac{\xi_1}{4\xi_3} \quad \textrm{and} \quad
\frac{k_{s2}}{k_{s1}}=\frac{-\xi_2}{1-\xi_3+3\xi_2} 
\end{equation}
and
\begin{equation}\label{relation1}
k_{s1}=4A\left(\frac{1-\xi_3+3\xi_2}{1-\xi_3-3\xi_2}\right)\frac{1}{\xi_1+2\xi_2},
\end{equation}
\begin{equation}\label{relation2}
k_{s2}=-4A\left(\frac{\xi_2}{1-\xi_3-3\xi_2}\right)\frac{1}{\xi_1+2\xi_2}
\end{equation}
and
\begin{equation}\label{relation3}
k_b^\prime = A\frac{\xi_1}{\xi_3}\left(\frac{1-\xi_3+3\xi_2}{1-\xi_3-3\xi_2}\right)\frac{1}{\xi_1+2\xi_2}.
\end{equation}

\subsection{The original stick and spiral model}
The corresponding results for the original stick and spiral model (i.e.
not including second nearest neighbor interactions for stretching) can
be obtained by setting $k_{s2}=0$. Thus, the solution of Eqs.~(\ref{length})
and (\ref{theta}) have again the form of (\ref{l_lambda}), with $\lambda_i$
and $\mu_{ij}$ given again by Eqs.~(\ref{lambda_i}) and (\ref{mu_ij}), but now
\begin{equation}
\xi_1=\frac{8k_b^\prime}{k_{s1}+18k_b^\prime}, \quad \xi_2=0 \quad 
\textrm{and} \quad \xi_3=\frac{2k_{s1}}{k_{s1}+18k_b^\prime}.
\end{equation}
The first of the Eqs.~(\ref{xi_dependence_2}) becomes $9\xi_1+2\xi_3=4$, while
the second remains the same.
The energy and the Young's modulus are again given by (\ref{gr_energy})
and (\ref{gr_Young}), respectively, but now
\begin{equation}
A=2k_{s1}k_b^\prime/(k_{s1}+18k_b^\prime),
\end{equation}
and the Poisson's ratio is
\begin{equation}
\nu=(2\xi_3-3\xi_1)/4=(k_{s1}-6k_{s2})/(k_{s1}+18k_b^\prime).
\end{equation}

Moreover, the relations between $k_{s1}$ and $k_b^\prime$, with $\xi_1$,
$\xi_3$ and $A$ are
\begin{equation}\label{orig_stick_spiral}
k_{s1}=4A/\xi_1 \quad \textrm{and}\quad k_b^\prime = A/\xi_3.
\end{equation}

\section{Force constants from DFT results and discussion}
\subsection{Details of our DFT calculations}

For our DFT calculations we used the Quantum 
Espresso \cite{quantesp} code at the level of GGA/PBE 
functional \cite{functional} and adopted an ultra-soft pseudopotential
for Carbon \cite{rrkj,pseudopotential}. 
The two unit cells are shown in Fig.~\ref{Fig4}. 
For the rectangular unit cell of Fig.~\ref{Fig4}(a) we used a 12$\times$12 
k-point mesh, while for the unit cell of Fig.~\ref{Fig4}(b) a 12$\times$6 (12 along the 
small real space direction).  In addition, we used cut-offs 50 and
500~Ryd for the wave functions and charge density, respectively, and
occupation smearing of 5~mRyd. As in 
Ref.~\onlinecite{PCCP_Fthenakis}, for non zero 
uniaxial strain, the unit cells were extended in the strain direction while all the atoms
in the cell as well as the vertical cell dimension were fully relaxed. 

\subsection{Results}

\begin{table*}
{\setlength{\tabcolsep}{0.3cm}
\begin{tabular}{@{}lcccccccc}
\hline
$n$ & $m$ & $\theta_0$ ($^o$) & $i$ & $\theta_{0i}$ ($^o$) & $\cos^2\theta_{0i}$ & $\lambda_i$ & $\cos 2\theta_{0i^\prime}$ & $\mu_{j^\prime k^\prime}$ \\
\hline\hline
1   & -1  & \ 90.000000 & 3 & 270.000000 & 0.000000        & -0.001556   & -1.000000       & \ 1.315279 \\
4   & -5  & 100.893395 & 3 & 259.106605 & 0.035714        & \ 0.008633  & -0.928571     &  \ 1.221761 \\
2   & \ 1  & \ 10.893395 & 1 & 109.106605 & 0.107143        & \ 0.027796   & -0.785714     & \ 1.032964 \\
1   & \ 1  & \ \ 0.000000  & 1, 2 & 120.000000 & 0.250000      & \ 0.066506    & -0.500000 & \ 0.654704 \\
2   & \ 1  & \ 10.893395 & 2 & 229.106605 & 0.428571        & \ 0.116258  & -0.142857 & \ 0.185116 \\
4   & -5  & 100.893395 & 2 & 139.106605 & 0.571429        & \ 0.156141  &  \ 0.142857 & -0.190542 \\
1   & -1  & \ 90.000000 & 1, 2 & \ 30.000000 & 0.750000      & \ 0.206426   &  \ 0.500000 & -0.657640 \\
4   & -5  & 100.893395 & 1 & \ 19.106605 & 0.892857         & \ 0.246905   &  \ 0.785714 & -1.031171 \\
2   & \ 1  & \ 10.893395 & 3 & 349.106605 & 0.964286        & \ 0.267113  &  \ 0.928571 & -1.218509 \\
1   & \ 1  & \ \ 0.000000  & 3 & 360.000000 & 1.000000        & \ 0.277621    &  \ 1.000000 & -1.309408 \\
\hline
\end{tabular}
\caption{\label{table1}Values of $\lambda_i$, $\mu_{ij}$ and $A$ obtained from the fittings for the four strain directions.}
}
\end{table*}

As a first step, we want to calculate the parameters $\lambda_i$ and
$\mu_{ij}$, which depend on the strain direction, as well as $A$, which
is independent. To calculate the $\lambda_i$ and
$\mu_{ij}$ values, we fit the
deformations $\delta l_i$ and $\delta \phi_{ij}$ in the strain range
$[-0.05, 0.05]$ to a quadratic form, considering that the coefficient 
of the linear term represent the corresponding $3a_0\lambda_i$ and
$\mu_{ij}$ values in Eq.~(\ref{l_lambda}), respectively. For the calculation
of $A$, we fit the
corresponding energy per atom values to a fourth order polynomial,
considering that $(3a_0)^2A$ is the coefficient of the quadratic term.

\begin{figure}
\begin{center}
\includegraphics[width=0.4\textwidth,clip]{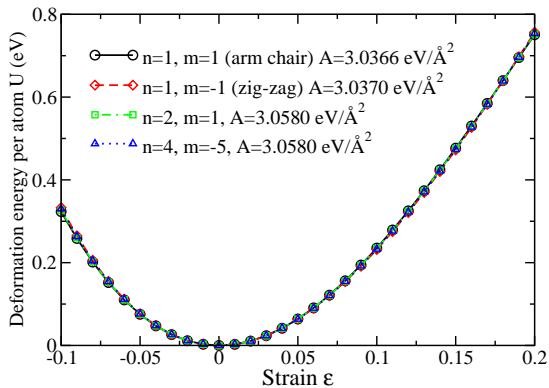}
\end{center}
\caption{(Color online)
Deformation energy per atom for strain along the direction
of the vectors $\mathbf{L}=n\mathbf{a}+m\mathbf{b}$, for $n$ and $m$
shown in the legends. For each strain direction, the $A$ values of 
(\ref{gr_energy}) are also presented in the legends.}
\label{energy}
\end{figure}

Although in real world, graphene sheet bends for negative strains,
computationally it is possible to perform calculations for negative
strains without bending of the structure. Fitting a curve to the deformations
$\delta l_i$, $\delta \phi_{ij}$ and $U$ for both negative and the positive
strain values, we expect a better estimation of
$\lambda_i$, $\mu_{ij}$ and $A$ values, than using an extrapolation of 
$\delta l_i$, $\delta \phi_{ij}$ and $U$ at $\varepsilon=0$, which can
be obtained from a fitting of the deformation
values of $\delta l_i$, $\delta \phi_{ij}$ and $U$
for positive strain values only.

Using the DFT method presented above, we calculated the deformations
$\delta l_i$ and $\delta\phi_{ij}$, $i,j=1,2,3$, and 
the deformation energy per atom $U$, for  uniaxial strain along the high symmetry
arm chair and zig-zag directions, as well as the directions along 
the vectors $\mathbf{L}=2\mathbf{a}+\mathbf{b}$ and
$\mathbf{L}_\perp=4\mathbf{a}-5\mathbf{b}$, which are perpendicular to
each other, and randomly selected. We increase the strain 
gradually with a 0.01 strain step in the range between $\varepsilon=-0.1$ and $\varepsilon=0.25$. The results are presented in Figs.~\ref{deformations}
and \ref{energy}, respectively.
The fitting functions are presented in the Supplementary Data. 

The values of $\lambda_i$ and $\mu_{ij}$ obtained from the fits for the
four strain directions are presented in Table~\ref{table1}, while the
corresponding $A$ values are shown in the legends of Fig.~\ref{energy}.
Although $A$ was expected to be independent of the strain direction, the
values of $A$ shown in Fig.~\ref{energy} does not seem to agree with this
prediction. However, this discrepancy is due to numerical errors introduced
from the different unit cells used. The total energy per atom
difference between the equilibrium graphene geometries at $\varepsilon=0$
obtained using the two unit cells of Fig.~\ref{Fig4} is 
$2.3\times 10^{-4}$~eV/atom. As one can show, this difference is enough to 
produce such a discrepancy in $A$, (i.e. of the order of 
$10^{-3}$~eV/\AA$^2$). It is worth noting, however, that the
difference between the two $A$ values, corresponding to the two
perpendicular strain directions of the same unit cell, is of the
order of $10^{-4}$~eV/\AA$^2$. For our calculations we will adopt the
value $A=3.046$~eV/\AA$^2$, which corresponds to an average of the
obtained values. 
 

\begin{figure*}
\begin{center}
\includegraphics[width=0.95\textwidth,clip]{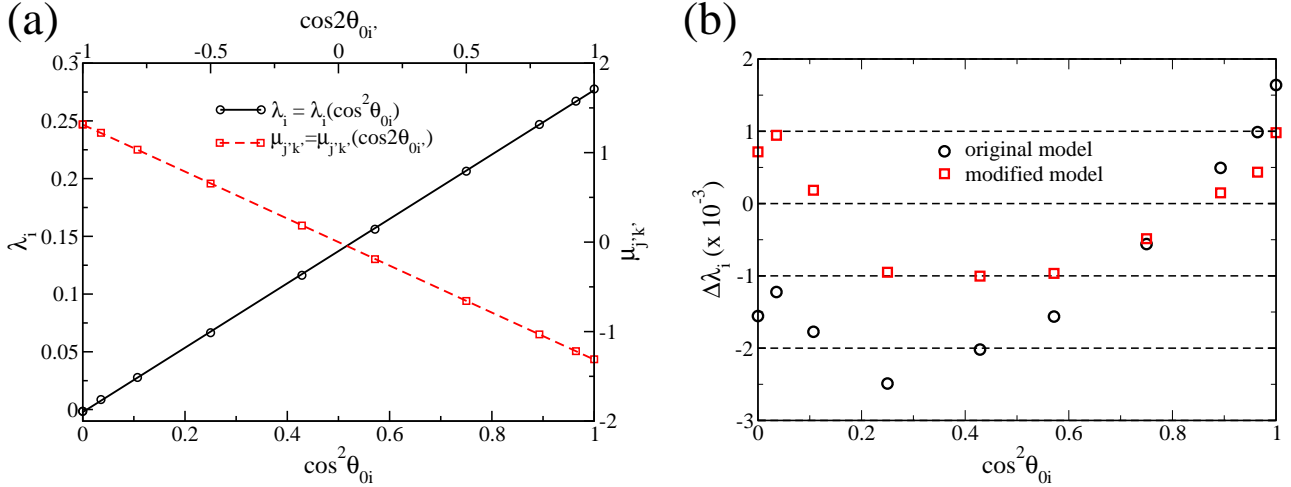}
\end{center}
\caption{(Color online)
(a) $\lambda_i$ and $\mu_{j^\prime k^\prime}$ as a 
function of $\cos^2\theta_{0i}$ and
$\cos 2\theta_{0i^\prime}$, respectively and the fitting lines, according to 
(\ref{lambda_i}) and (\ref{mu_ij_second}). (b) Difference $\Delta\lambda_i$ between
the values $\lambda_i$ of Table~\ref{table1} and those predicted by fitting equations
of $\lambda_i$ as a function of $\cos^2\theta_{0i}$. }
\label{lambda_mu_values}
\end{figure*}

The second step is to calculate the values of $\xi_1$, $\xi_2$ and $\xi_3$
using the $\lambda_i$ and $\mu_{ij}$ values of Table~\ref{table1} 
and Eqs.~(\ref{lambda_i}) and (\ref{mu_ij}). 
According to these equations, $\xi_1$, $\xi_2$
and $\xi_3$ can be obtained using a linear fitting
of the $\lambda_i$ values as a function of $\cos^2\theta_{0i}$ and the 
$-\mu_{i^\prime j^\prime}/\sqrt{3}$ values as a function of
$\cos 2\theta_{0k^\prime}$.
The values of $\lambda_i$ as a function of $\cos^2\theta_{0i}$ and the values of
$-\mu_{i^\prime j^\prime}/\sqrt{3}$ as a function of $\cos 2\theta_{0k^\prime}$,
as well as the corresponding fitting lines are shown in 
Fig.~\ref{lambda_mu_values}(a). The smoothness of the fitting is obvious.
These fitting lines are
\begin{equation}
\lambda_i = 0.278912\cos^2\theta_{0i}-0.002272
\end{equation}
and
\begin{equation}
\mu_{i^\prime j^\prime} = -0.758145\sqrt{3}\cos2\theta_{0k^\prime} .
\end{equation}
Thus, $\xi_1=0.278921$, $\xi_2=-0.002272$ and $\xi_3=0.758145$.
Using these values, the value of $A$, and Eqs.~(\ref{relations1})~-~(\ref{relation3}), 
we can calculate the values of $k_{s1}$, 
$k_{s2}$ and $k_b^\prime$, as well as the ratios $k_{s2}/k_{s1}$
and $k_b^\prime/k_{s1}$. Thus,
$k_b^\prime/k_{s1}=0.091975$, $k_{s2}/k_{s1}=0.0096665$, 
$k_{s1}=41.972$~eV/\AA, $k_{s2}=0.40572$~eV/\AA{} and
$k_b^\prime=3.8604$~eV/\AA.
Therefore, roughly speaking $k_b^\prime \approx 0.1 k_{s1}$ 
and $k_{s2}\approx 0.01 k_{s1}$,
which qualitatively provides the relative strength of each deformation mode.
Moreover, according to (\ref{gr_Young}) and (\ref{nu}), $E=1012$~GPa 
and $\nu=0.1744$, in agreement with the results of our previous work 
\cite{PCCP_Fthenakis} obtained fitting the stress $\sigma$ and the
the transverse strain $\varepsilon_\perp$ values as a function of strain,
to a third and second order polynomial, respectively.

Knowing the $k_{s1}$, $k_{s2}$ and $k_b^\prime$ values, we have the ability to
predict any mechanical property related to the in-plane deformations
of graphene and not only $E$ and $\nu$. For instance, the corresponding
biaxial isotropic modulus $E_B=\sigma/\varepsilon$, where 
$\sigma=\sigma_{xx}=\sigma_{yy}$
and $\varepsilon=\varepsilon_{xx}=\varepsilon_{yy}$, is 
$E_B=4\sqrt{3}A^\prime/d_0$, where for the biaxial isotropic deformation
$U=9a_0^2A^\prime \varepsilon^2$. Using (\ref{grgr1}) and 
(\ref{grgr2}), it is easy to show that for biaxial
isotropic strain $A^\prime=k_{s1}/6+k_{s2}$. Thus, for graphene, 
$E_B=2459$~GPa. A different calculation using the relation
$U=k\delta l^2/2+k(\delta l+\delta l_\perp)^2/2=k\delta l^2(1+\nu+\nu^2/2)=
2U_u(1+\nu+\nu^2/2)$, or $A^\prime=2(1+\nu+\nu^2/2)A$, yields
$E_B=2408$~GPa. As one can see, the two results are very close to each other.

Obviously, the term $U_2$ corresponding to
the stretching of the second nearest neighbor interatomic distances is the
less important energy contribution, but it is not a term that can be 
ignored. If this term is ignored, (which is equivalent to set $k_{s2}=0$
or $\xi_2=0$), the energy model reduces to the original
stick and spiral model, which, according to (\ref{length}), 
predicts that any bond which is
perpendicular to the strain direction remains undeformed. This,
 however, is in contrast
to what we find from our DFT calculations for the $l_3$ bond length 
under uniaxial strain along the zig-zag direction. 
Just for comparison, we also calculate the corresponding $\xi_1$, $\xi_3$, $k_{s1}$
and $k_b^\prime$ values obtained from the original stick and spiral morel.
Obviously, the form of Eq.~(\ref{mu_ij}) does not change in the original stick
and spiral model and consequently the value of $\xi_3$ remains the same
as the modified model. However, (\ref{lambda_i}) becomes $\lambda_i=\xi_1\cos^2\theta_{0i}$.
The corresponding fit for the $\lambda_i$ values of Table~\ref{table1} as a function
of $\cos^2\theta_{0i}$ yelds $\xi_1=0.275981$. In Fig.~\ref{lambda_mu_values}(b)
we show the prediction error $\delta \lambda_i$ (i.e. the difference between the
$\lambda_i$ provided by the fitting equations of $\lambda_i$ as a function of 
$\cos^2\theta_{0i}$ and the corresponding $\lambda_i$ values of Table~\ref{table1} for
the original and the modified stick and spiral model. As we can see, the error for
the modified sick and spiral model is between $\pm 0.001$, while the error for the 
original model is almost double, ranging between -0.0025 and 0.0017. The values of 
$k_{s1}$ and $k_{b}^\prime$ for the original model, according to (\ref{orig_stick_spiral}) 
are $k_{s1}=44.178$~eV/\AA{}
and $k_b^\prime=4.0177$~eV/\AA, i.e. they are overesimated by 5 and 4\%, respectively,
in comparisson with the corresponding values obtained from the modified model.
Thus, the original stick and spiral model can not provide an accurate
description for the bond and angle deformations of graphene, or at least,
it can not provide such an accurate description as 
the modified model, which is presented here.

\section{Conclusions}
In summary, we present a method for the study of the equilibrium deformations 
of 2D planar materials under uniaxial strain. The method is based on the stick
and spiral model including angle bending energy terms and either only 1st
nearest neighbors bond stretching terms (case 1) or both  1st and 2nd nearest
neighbors terms (case 2). The method can be generalized to describe structural
deformations not only under uniaxial strain, but also under any strain
conditions. We present analytic expressions/equations for the structure
deformations under strain, namely the equilibrium angle bending and bond
stretching deformations for both case 1 (equations (\ref{length}) and 
(\ref{theta})) and case 2 (equations (\ref{length2}) and (\ref{theta2})). 
We then focus on graphene in order to assess the applicability of our method for which we perform
DFT calculations for several values of strain in 4 different directions. We find that the original stick and spiral model (case 1) decouples the equations yielding 
$\delta l_i$ from those yielding $\delta \theta_i$ and for graphene, it predicts that the vertical to 
the strain bonds are not modified. This is in contrast with the DFT results. The
inclusion of 2nd nearest neighbors stretching terms (case 2) results in the coupling
of $\delta l_i$ and $\delta \theta_i$, improves the model significantly 
and brings the results in close agreement with DFT. Our method provides
a simple and solid method to study the structural deformations of Graphene in
the case of uniaxial strain on any direction in the elastic regime. The elastic
properties of graphene under strain are very accurately reproduced by  our method.
Although this first application concerns graphene, our method can be applied to
any 2D planar material and it would be interesting to assess its accuracy on
different structures and materials like Graphene planar allotropes, h-BN,
Si$_3$B, Si$_2$BN, CdS, etc.  

\section*{Acknowledgements}
NNL acknowledges support from the Hellenic Ministry of
Education (through ESPA) and from the GSRT through ``Advanced Materials and Devices'' program (MIS:5002409).

\appendix
\section{Relation between $\phi_{ij}$ and $\theta_i$s}\label{delta_phi_delta_theta}
Let us define, for each atom of the unit cell, a local anti-clockwise
frame of coordinates with its origin at the position of that atom and
its x-axis along the strain direction, as shown in Fig.~\ref{Fig1}(c). 
Let us denote as $\mathbf{r}_{1}$, $\mathbf{r}_{2}$ and $\mathbf{r}_{3}$
the three bond vectors, which have their tail on atom $i$ and 
by $\theta_{1}$, $\theta_{2}$ and $\theta_{3}$ the corresponding
angles between these bond vectors with the strain direction, respectively,
as shown in Fig.~\ref{Fig1}(c).

Obviously, $\mathbf{r_i}\mathbf{r_j}=r_i r_j \cos\phi_{ij}$, where $\phi_{ij}$ is
the angle formed by the bonds $i$ and $j$, and 
$\mathbf{r}_i=r_i \cos\theta_i \hat{\mathbf{i}}+r_i\sin\theta_i \hat{\mathbf{j}}$, $i=1,2,3$.
Thus, the dot product $\mathbf{r_i}\mathbf{r_j}$ can be written as
\begin{eqnarray}
\mathbf{r_i}\mathbf{r_j} & = & (r_i \cos\theta_i \hat{\mathbf{i}}+r_i\sin\theta_i \hat{\mathbf{j}})
(r_j \cos\theta_j \hat{\mathbf{i}}+r_j\sin\theta_j \hat{\mathbf{j}}) \nonumber \\
 & = & r_i r_j \cos(\theta_j-\theta_i),
\end{eqnarray}
and consequently, 
\begin{equation}\label{cos}
\cos \phi_{ij}=\cos(\theta_j-\theta_i)
\end{equation}
If $\phi_{0ij}$, $\theta_{0i}$ and $\theta_{0j}$ are the values of the corresponding 
$\phi_{ij}$, $\theta_{i}$ and $\theta_{j}$ angles at equilibrium for 
$\varepsilon=0$, then using a first order Taylor expansion around these values, 
Eq.~(\ref{cos}) yields 
\begin{equation}\label{sin}
\sin \phi_{0ij}\delta\phi_{ij}=\sin(\theta_{0j}-\theta_{0i})(\delta\theta_j-\delta\theta_i),
\end{equation}
where $\phi_{ij}=\phi_{0ij}+\delta\phi_{ij}$, $\theta_i=\theta_{0i}+\delta\theta_i$ and
$\theta_j=\theta_{0j}+\delta\theta_j$ are the corresponding angles at $\varepsilon\ne 0$.
Thus, the derivative of $\delta\phi_{ij}$ with respect to $\delta\theta_i$ is
\begin{equation}\label{partial_phi}
\partial\delta\phi_{ij}/\partial\delta\theta_i=\sin(\theta_{0i}-\theta_{0j})/\sin\phi_{0ij}.
\end{equation}

Imposing that $0 < \phi_{ij}\le\pi$, (\ref{cos}) gives
\begin{equation}\label{A3}
-2k\pi < \pm |\theta_i-\theta_j| \le (1-2k)\pi.
\end{equation}
If $\theta_i$s, $i=1,2,3$ are defined inside the same unit circle (e.g. 
$0\le \theta_i < 2\pi$ or $-\pi<\theta_i\le\pi$), then 
$-2\pi<\theta_i-\theta_j<2\pi$.
However, according to (\ref{A3}), $\theta_i-\theta_j$ is out of the
range $(-2\pi, 2\pi)$, for $k \neq 0$ or $1$, and therefore only $k=0$
and $k=1$ should be considered. Consequently, (i) for $k=0$ 
(or $0<|\theta_i-\theta_j| \le \pi$, according to (\ref{A3})), 
$\phi_{ij}=|\theta_i-\theta_j|$ and (ii) for $k=1$ 
(or $\pi \le |\theta_i-\theta_j| < 2\pi$, according to (\ref{A3})),
$\phi_{ij}=2\pi-|\theta_i-\theta_j|$. 
Thus, for any case, $\delta\phi_{ij}=\pm(\delta\theta_i-\delta\theta_j)$,
which leads to (\ref{square_delta}).

If $\mathbf{r}_i$s, $i=1,2,3$, have their tail at the position of an atom A, then they have their head at the position of the 
atoms which form bonds with atom A. Assume B is such an atom, which forms a bond with another atom C (different than A), and
$\mathbf{r}_1$ and $\mathbf{r}_4$ are the bond vectors corresponding to the bonds A-B and B-C, respectively. 
There are two options for the direction of $\mathbf{r}_4$: either
its head is on the position of atom B and its tail on the position of atom C, or the opposite. In the former case, the relations
between the bond angle $\phi_{ij}$ and the bond angle $\theta_i$ with respect to the strain direction are the same with those
presented above, since $\mathbf{r}_1\mathbf{r}_4=r_1r_4\cos\phi_{14}$. However, in the later case, 
$\mathbf{r}_1\mathbf{r}_4=r_1r_4\cos\omega_{14}$, where the bond angle $\phi_{14}$ is $\phi_{14}=\pi-\omega_{14}$. Thus, for
this case, the relations presented above will be valid if $\phi_{ij}$ is replaced by $\pi-\phi_{ij}$. Thus, (\ref{cos}),
should be replaced by 
\begin{equation}\label{cos_inv}
\cos\phi_{ij}=-\cos(\theta_i-\theta_j),
\end{equation}
\begin{equation}\label{sin_inv}
\sin \phi_{0ij}\delta\phi_{ij}=-\sin(\theta_{0j}-\theta_{0i})(\delta\theta_j-\delta\theta_i),
\end{equation}
and 
\begin{equation}\label{partial_phi_inv}
\partial\delta\phi_{ij}/\partial\delta\theta_i=-\sin(\theta_{0i}-\theta_{0j})/\sin\phi_{0ij}.
\end{equation}

If $0\le\pi-\phi_{ij}<\pi$, then $0<\phi_{ij}\le\pi$. For $\phi_{ij}$ in this range, (\ref{cos_inv}) yields
(i) if $0<|\theta_i-\theta_j|\le\pi$, then $\phi_{ij}=\pi-|\theta_i-\theta_j|$ and (ii)
if $\pi<|\theta_i-\theta_j|\le 2\pi$, then $\phi_{ij}=|\theta_i-\theta_j|-\pi$.
Obviously, therefore, for this case, $\delta\phi_{ij}$ is also $\delta\phi_{ij}=\pm(\delta\theta_i-\delta\theta_j)$ and 
consequently, (\ref{square_delta}) is also valid.

\section{The physical meaning of $\lambda$}\label{Physical_meaning}
Obviously, $\Lambda$ is parametrically dependent on $\varepsilon$, i.e.
$\Lambda=\Lambda(\{\delta l_i\},\{\delta\theta_{i}\},\lambda;\varepsilon)$.
If $\Lambda$ is minimized for $\delta l_i=\delta l_i^*$, $\delta \theta_i=\delta\theta_i^*$, 
and $\lambda=\lambda^*$, where $\delta l_i^*$s, $\delta\theta_i^*$s and $\lambda^*$ are
specific values of $\delta l_i$s, $\delta \theta_i$s and $\lambda$, respectively,
then $\Lambda_{min}=
\Lambda(\{\delta l_i^*\},\{\delta\theta_{i}^*\},\lambda^*;\varepsilon)=\Lambda_{min}(\varepsilon)$,
where $\Lambda_{min}$ is the minimum of $\Lambda$.

For $\delta l_i=\delta l_i^*$ and $\delta \theta_i=\delta\theta_i^*$,
the strain $\varepsilon$ is
$\varepsilon=\delta L(\{\delta l_i^*\},\{\delta\theta_{i}^*\})/L_0$ and 
$U$ is minimized subject to the constrain
$\varepsilon=\delta L/L_0$. Thus,
if $U_{min}$ is the minimum of $U$ 
subject to the constrain $\varepsilon=\delta L/L_0$, then
$U_{min}=U(\{\delta l_i^*\},\{\delta\theta_{i}^*\})$ and 
(according to (\ref{constrain})), $U(\{\delta l_i^*\},\{\delta\theta_{i}^*\})=
\Lambda(\{\delta l_i^*\},\{\delta\theta_{i}^*\},\lambda^*;\varepsilon)$, or
$U_{min}(\varepsilon)=\Lambda_{min}(\varepsilon)$.

According to (\ref{length}) and (\ref{theta}), for the minimized $\Lambda$,
$\delta l_i^*$ and $\delta \theta_i^*$ depend linearly on $\lambda^*$, and therefore, according to 
(\ref{constrain}), $\lambda^*$ should depend linearly on $\varepsilon$. Thus,
$\delta l_i^*=\delta l_i^*(\varepsilon)$ and $\delta \theta_i^*=\delta \theta_i^*(\varepsilon)$,
and consequently, $U_{min}=U_{min}(\varepsilon)$.
On the other hand, $U_{min}$ is quadratically dependent on $\delta l_i^*$ and $\delta \theta_i^*$, and consequently
$U_{min}$ should depend quadratically on $\varepsilon$. Therefore we can write $U_{min}(\varepsilon)=K\varepsilon^2$,
where $K=K(\{k_{si}\},\{k_{bij}\})$.

Obviously, $\partial \Lambda(\{\delta l_i\},\{\delta\theta_i\},\lambda;\varepsilon)/\partial\varepsilon=\lambda$,
and consequently, $d\Lambda_{min}/d\varepsilon=d\Lambda(\{\delta l_i^*\},\{\delta\theta_i^*\},\lambda^*;\varepsilon)/d\varepsilon=\lambda^*$.
Since, $\Lambda_{min}=U_{min}$, we have $d\Lambda_{min}/d\varepsilon=dU_{min}/d\varepsilon=2K\varepsilon$.
Thus, $2K\varepsilon=\lambda^*$, which leads to (\ref{U_min}).

\section{$\delta r_{ij}$ as a function of bond length and bond angle deformations}\label{delta_r_ij}

Let us assume that atoms A, B and C belong to the same planar 2D structure
and atom A forms bonds with atoms B and C. 
Let us also assume that $\mathbf{r}_{0i}$ and $\mathbf{r}_{0j}$ are the bond vectors
corresponding to the bonds A-B and A-C at equilibrium for 
$\varepsilon=0$, having both their tails (or their heads) at the 
position of atom A. Then the interatomic distance $r_{0ij}$ between atoms B and C
is the length of the vector $\mathbf{r}_{0ij}=\mathbf{r}_{0j}-\mathbf{r}_{0i}$,
for which
\begin{equation}
r_{0ij}^2 = l_{0i}^2+l_{0j}^2-2l_{0i}l_{0j}\cos\phi_{0ij}.
\end{equation}
where $l_{0i}$ and $l_{0j}$ are the lengths of $\mathbf{r}_{0i}$ and
$\mathbf{r}_{0j}$, respectively, and $\phi_{0ij}$ the bond angle between
bonds A-B and A-C.
If at the equilibrium state under strain, $l_{0i}$, $l_{0j}$, $r_{0ij}$
and $\phi_{0ij}$ are deformed to $l_i=l_{0i}+\delta l_i$,
$l_j=l_{0j}+\delta l_j$, $r_{ij}=r_{0ij}+\delta r_{ij}$ and
$\phi_{ij}=\phi_{0ij}+\delta\phi_{ij}$, respectively, then
\begin{eqnarray}\label{C2}
r_{ij}^2 & = & (l_{0i}+\delta l_i)^2+(l_{0j}+\delta l_j)^2 \nonumber \\
         &   & -2(l_{0i}+\delta l_i)(l_{0j}+\delta l_j)\cos(\phi_{0ij}+\delta\phi_{ij}) \nonumber \\
         &\approx & l_{0i}^2+2l_{0i}\delta l_i+l_{0j}^2+2l_{0j}\delta l_j \nonumber \\
         &   & -2(l_{0i}l_{0j}+l_{0i}\delta l_j+
               l_{0j}\delta l_i)(\cos\phi_{0ij}-\sin\phi_{0ij}\delta\phi_{ij}) \nonumber \\
         &\approx & r_{0ij}^2 +2(l_{0i}\delta l_i+l_{0j}\delta l_j-l_{0i}\cos\phi_{0ij}\delta l_j \nonumber \\
         &   & -l_{0j}\cos\phi_{0ij}\delta l_i+l_{0i} l_{0j}\sin\phi_{0ij}\delta\phi_{ij}).
\end{eqnarray}
For $\delta r_{ij}<<r_{0ij}$,
$r_{ij}^2 \approx r_{0ij}^2+2r_{0ij}\delta r_{ij},$ 
and consequently, (\ref{C2}) leads to
\begin{eqnarray}\label{second}
r_{0ij}\delta r_{ij} & = & \left(l_{0i}-l_{0j}\cos\phi_{0ij}\right)\delta l_i
                     +\left(l_{0j}-l_{0i}\cos\phi_{0ij}\right)\delta l_j \nonumber \\
                     &   & +l_{0i} l_{0j}\sin\phi_{0ij}\delta\phi_{ij}.
\end{eqnarray}
Therefore, $\delta r_{ij}$ is a function of the deformations of $\delta l_i$,
$\delta l_j$, $\delta\theta_i$ and $\delta\theta_j$, (see 
Sec.~\ref{delta_phi_delta_theta}).

The derivatives of $\delta r_{ij}$ with respect to $\delta l_i$ and 
$\delta \theta_i$ are
\begin{equation}\label{partial_l}
\partial\delta r_{ij}/\partial\delta l_i = \left[l_{0i}-l_{0j}\cos\phi_{0ij}\right]/r_{0ij}
\end{equation}
and
\begin{equation}\label{partial_theta}
\partial\delta r_{ij}/\partial\delta\theta_i = \left[l_{0i} l_{0j}\sin\phi_{0ij}/r_{0ij}\right](\partial\delta\phi_{ij}/\partial\delta\theta_i).
\end{equation}

Using (\ref{cos}), (\ref{sin}) and (\ref{partial_phi}) the above equations give
\begin{eqnarray}\label{second_1}
r_{0ij}\delta r_{ij} & = & \left(l_{0i}-l_{0j}\cos(\theta_{0i}-\theta_{0j})\right)\delta l_i \nonumber \\
                     &   & +\left(l_{0j}-l_{0i}\cos(\theta_{0j}-\theta_{0i})\right)\delta l_j \nonumber \\
                     &   & +l_{0i} l_{0j}\sin(\theta_{0i}-\theta_{0j})(\delta\theta_{i}-\delta\theta_{j}),
\end{eqnarray}
and
\begin{equation}\label{partial_l_1}
\partial\delta r_{ij}/\partial\delta l_i = \left[l_{0i}-l_{0j}\cos(\theta_{0i}-\theta_{0j})\right]/r_{0ij},
\end{equation}
\begin{equation}\label{partial_theta_1}
\partial\delta r_{ij}/\partial\delta\theta_i = l_{0i} l_{0j}\sin(\theta_{0i}-\theta_{0j})/r_{0ij}.
\end{equation}

However, if the head of $\mathbf{r}_i$ and the tail of $\mathbf{r}_j$ (or vice versa) 
are at the position of atom A, then we have to use (\ref{cos_inv}), (\ref{sin_inv})
and (\ref{partial_phi_inv}) instead of (\ref{cos}), (\ref{sin}) and 
(\ref{partial_phi}) (see Sec.~\ref{delta_phi_delta_theta}), and thus,
(\ref{second}), (\ref{partial_l}) and (\ref{partial_theta}) give
\begin{eqnarray}\label{second_2}
r_{0ij}\delta r_{ij} & = & \left(l_{0i}+l_{0j}\cos(\theta_{0i}-\theta_{0j})\right)\delta l_i \nonumber \\
                     &   & +\left(l_{0j}+l_{0i}\cos(\theta_{0j}-\theta_{0i})\right)\delta l_j \nonumber \\
                     &   & -l_{0i} l_{0j}\sin(\theta_{0i}-\theta_{0j})(\delta\theta_{i}-\delta\theta_{j}),
\end{eqnarray}
and
\begin{equation}\label{partial_l_2}
\partial\delta r_{ij}/\partial\delta l_i = \left[l_{0i}+l_{0j}\cos(\theta_{0i}-\theta_{0j})\right]/r_{0ij},
\end{equation}
\begin{equation}\label{partial_theta_2}
\partial\delta r_{ij}/\partial\delta\theta_i = -l_{0i} l_{0j}\sin(\theta_{0i}-\theta_{0j})/r_{0ij}.
\end{equation}
Commuting $i$ with $j$ in (\ref{partial_l}), (\ref{partial_theta}), 
(\ref{partial_l_1}), (\ref{partial_theta_1}), (\ref{partial_l_2}) and 
(\ref{partial_theta_2}), we obtain
the corresponding relations for $\partial\delta r_{ij}/\partial\delta\theta_j$ and
$\partial\delta r_{ij}/\partial\delta l_j$.

\section{Derivation of Eqs.~(\ref{q_s}), (\ref{gr_strain2}) and (\ref{normal_proj_gr})}\label{graphene_strain}
If $\mathbf{L}_0=n\mathbf{a}+m\mathbf{b}$ defines the strain direction, then
$\mathbf{L}_0=(\sqrt{3}/2)(\sqrt{3}(n+m)\hat{{\mathbf{i}}}+(n-m)\hat{\mathbf{j}})a_0$,
and consequently, 
$\cos\theta_0 = 3(n+m)a_0/(2L_0)$ and $\sin\theta_0=\sqrt{3}(n-m)a_0/(2L_0)$,
where $\theta_0$ is the angle of the strain direction with respect to the
x-axis. Solving these two equations with respect to $n$ and $m$, we obtain, 
$n=2L_0/(3a_0)((1/2)\cos\theta_0+(\sqrt{3}/2)\sin\theta_0)=-2L_0/(3a_0)\cos\theta_{02}$
and 
$m=2L_0/(3a_0)((1/2)\cos\theta_0-(\sqrt{3}/2)\sin\theta_0)=-2L_0/(3a_0)\cos\theta_{01}$,
and consequently,
$n+m=2L_0/(3a_0)\cos\theta_0=2L_0/(3a_0)\cos\theta_{03}$, which lead to
(\ref{q_s}). In Sec.~\ref{useful} we present useful relations
between the trigonometric functions of these angles, which will be used here.

Bearing in mind that in graphene $l_{01}=l_{02}=l_{03}=a_0$, and using
(\ref{q_s}), (\ref{constrain}) becomes 
\begin{equation}
\varepsilon = \frac{2}{3a_0}\sum_{i=1}^3\cos^2\theta_{0i}\delta l_i- 
         \frac{1}{3}\sum_{i=1}^3\sin2\theta_{0i}\delta\theta_i.
\end{equation}
Using (\ref{sin1}) for $k=2$ of Sec.~\ref{useful}, the above
equation leads to Eq.~(\ref{gr_strain2}). 

Moreover, (\ref{normal_proj}) becomes
\begin{equation}
\begin{aligned}
& \sum_j \cos\theta_{0j}(\delta l_j\sin\theta_{0j}+a_0\cos\theta_{0j}\delta\theta_j) = 0 \Rightarrow  \\
& \sum_j(\delta l_j\sin 2\theta_{0j}/2+a_0\cos^2\theta_{0j}\delta\theta_j) = 0 \Rightarrow  \\
& \sum_j(\delta l_j\sin 2\theta_{0j}/2+a_0\cos^2\theta_{0j}(\delta\theta_j-\delta\theta_i))=  \\
&\qquad  = -a_0\delta\theta_i\sum_j\cos^2\theta_{0j} \nonumber \Rightarrow \\
& \sum_j(\delta l_j\sin 2\theta_{0j}+2a_0\cos^2\theta_{0j}(\delta\theta_j-\delta\theta_i)) = -3a_0\delta\theta_i,
\end{aligned}
\end{equation}
which leads to (\ref{normal_proj_gr}).
In the last step of the above equation we used (\ref{cos2}) of the Sec.~\ref{useful}.

\section{Derivation of Eqs.~(\ref{gr_2_length}), (\ref{gr_2_angle}), (\ref{delta_l_i}), 
(\ref{delta_theta_ij}) and (\ref{xi_dependence})}\label{modified}
As we can see in Fig.~\ref{graphene_unit}, the tails of the bond
vectors $\mathbf{r}_1$, $\mathbf{r}_2$ and $\mathbf{r}_3$ are at the position of atom A, while
the heads of the bond vectors $\mathbf{r}_4$, $\mathbf{r}_5$ and $\mathbf{r}_6$ are at the
position of atom B. Therefore, to apply (\ref{second_fin}), (\ref{partial_l_fin})
and (\ref{partial_theta_fin}) to (\ref{length2})
and (\ref{theta2}), we have to use the upper signs among $\pm$ and $\mp$.
Moreover, $l_{0i}=a_0$, $r_{0ij}=\sqrt{3}a_0$,
$\cos(\theta_{0j^\prime}-\theta_{0i^\prime})=\cos(2\pi/3)=-1/2$ and 
$\sin(\theta_{0j^\prime}-\theta_{0i^\prime})=\sin(2\pi/3)=\sqrt{3}/2$.
Consequently, (\ref{second_fin}), (\ref{partial_l_fin}) and 
(\ref{partial_theta_fin}) yield
\begin{equation}
\delta r_{i^\prime,j^\prime} = (\sqrt{3}/2)(\delta l_{i^\prime}+\delta l_{j^\prime})+(a_0/2)(\delta\theta_{j^\prime}-\delta\theta_{i^\prime}),
\end{equation}
$\partial \delta r_{ij}/\partial\delta l_i = \sqrt{3}/2$ and $\partial \delta r_{i^\prime j^\prime}/\partial \delta\theta_{i^\prime} = -\partial \delta r_{i^\prime j^\prime}/\partial \delta \theta_{j^\prime} = -a_0/2$, respectively.
Thus, (\ref{length2}) gives 
\begin{equation}
\begin{aligned}
& k_{s1}\delta l_i+\frac{\sqrt{3}}{2}k_{s2}\sum_{k=1}^4\delta r_{ij_k}=\frac{\lambda q_i \cos\theta_{0i}}{L_0} \Rightarrow \nonumber \\
& k_{s1}\delta l_i+\frac{\sqrt{3}}{2}k_{s2}2(\delta r_{ij}+\delta r_{ki})=\frac{2L_0\cos\theta_{0i}}{3a_0}\frac{\lambda\cos\theta_{0i}}{L_0} \Rightarrow \nonumber \\
& k_{s1}\delta l_{i^\prime}+\sqrt{3}k_{s2}\left[\frac{\sqrt{3}}{2}(\delta l_{i^\prime}+\delta l_{j^\prime})+\frac{a_0}{2}(\delta \theta_{j^\prime}-\delta\theta_{i^\prime})+ \right. \\
& \qquad \left.\frac{\sqrt{3}}{2}(\delta l_{k^\prime}+\delta l_{i^\prime})+\frac{a_0}{2}(\delta \theta_{i^\prime}-\delta\theta_{k^\prime})\right]=\frac{2\lambda}{3a_0}\cos^2\theta_{0i^\prime}, \nonumber
\end{aligned}
\end{equation}
which leads to (\ref{gr_2_length}), and (\ref{theta2}) gives
\begin{equation}
\begin{aligned}
& 2k_b^\prime a_0^2[(\delta\theta_i-\delta\theta_j)+(\delta\theta_i-\delta\theta_k)]+ \\
& \qquad 2k_{s2}[\delta r_{ij}(\partial\delta r_{ij}/\partial\delta\theta_i) +
\delta r_{ki}(\partial\delta r_{ki}/\partial\delta\theta_i)]= \\
& \qquad =-\lambda q_i a_0 \sin\theta_{0i}/L_0 \Rightarrow \\
& 2k_b^\prime a_0^2 [(\delta\theta_i-\delta\theta_j)+(\delta\theta_i-\delta\theta_k)]+ a_0k_{s2}(\delta r_{ki}-\delta r_{ij}) = \\
&\qquad =-\lambda (2L_0\cos\theta_{0i})/(3a_0) a_0 \sin\theta_{0i}/L_0 \Rightarrow \\
& k_b^\prime a_0^2 [(\delta\theta_{i^\prime}-\delta\theta_{j^\prime})+(\delta\theta_{i^\prime}-\delta\theta_{k^\prime})]+ \\
& \qquad a_0k_{s2}/2[(\sqrt{3}/2)(\delta l_{k^\prime}+\delta l_{i^\prime})+(a_0/2)(\delta\theta_{i^\prime}-\delta\theta_{k^\prime})- \\
& \qquad (\sqrt{3}/2)(\delta l_{i^\prime}+\delta l_{j^\prime})-(a_0/2)(\delta\theta_{j^\prime}-\delta\theta_{i^\prime})] = \\
& \qquad = -(\lambda/6)\sin 2\theta_{0i^\prime},
\end{aligned}
\end{equation}
which leads to (\ref{gr_2_angle}).
Summing up the three equations (\ref{gr_2_length}) (i.e. for 
$(i^\prime,j^\prime,k^\prime)=(1,2,3)$,
$(2,3,1)$ and $(3,1,2)$), and using (\ref{cos2}) we obtain
\begin{equation}\label{gr_l_sum}
(k_{s1}+6k_{s2})(\delta l_1+\delta l_2+\delta l_3)=\lambda/a_0.
\end{equation}\
Substituting $\sum_{i=1}^3\delta l_i$ in (\ref{gr_2_length}) we take
\begin{equation}\label{gr_2_length2}
\begin{aligned}
& \left(k_{s1}+3k_{s2}/2\right)\delta l_{i^\prime}+(\sqrt{3}/2)k_{s2}a_0(\delta\theta_{j^\prime}-\delta\theta_{k^\prime}) \\
& = (\lambda/a_0)\left[(2/3)\cos^2\theta_{0i^\prime}-(3/2)k_{s2}/(k_{s1}+6k_{s2})\right]. 
\end{aligned}
\end{equation}
Subtracting by parts equations (\ref{gr_2_angle}) (two at a time) leads to
\begin{eqnarray}\label{gr_2_theta2}
& & 3\left(k_b^\prime+k_{s2}/4\right)a_0^2(\delta\theta_{j^\prime}-\delta\theta_{k^\prime})+ \nonumber \\
& & (\sqrt{3}/4)a_0k_{s2}(3\delta l_{i^\prime}-(\delta l_1+\delta l_2+\delta l_3)) \nonumber \\
& = & (\lambda/6)(\sin2\theta_{0k^\prime}-\sin2\theta_{0j^\prime}) \Longleftrightarrow \nonumber \\
& & (\sqrt{3}/4)a_0k_{s2}\delta l_{i^\prime} + \left(k_b^\prime+k_{s2}/4\right)a_0^2(\delta\theta_{j^\prime}-\delta\theta_{k^\prime}) \nonumber \\
& = & \frac{\lambda}{2\sqrt{3}}\left[\frac{2}{3}\cos^2\theta_{0i^\prime}-\frac{1}{3}+\frac{k_{s2}}{2(k_{s1}+6k_{s2})}\right].
\end{eqnarray}
The solution of the system of (\ref{gr_2_length2}) and 
(\ref{gr_2_theta2}) are (\ref{delta_l_i}) and (\ref{delta_theta_ij}).

Using the expressions of (\ref{delta_l_i}) and (\ref{delta_theta_ij})
for $\delta l_i$ and $\delta \theta_j - \delta\theta_i$, and (\ref{sin1}),
(\ref{cos2}), (\ref{sin2}), (\ref{cos4}) and (\ref{cos3_sin1}), 
(\ref{gr_strain2}) and (\ref{normal_proj_gr}) become
\begin{eqnarray}
\varepsilon & = & \frac{2}{3a_0}\sum_{i=1}^3\cos^2\theta_{0i}3a_0(\xi_1^\prime\cos^2\theta_{0i}+\xi_2^\prime) \nonumber \\
& & -\frac{1}{3}\sum_{i=1}^3\sin2\theta_{0i}\xi_3^\prime(\sin2\theta_{0j}-\sin2\theta_{0i}) \nonumber \\
& = & 2\left[\xi_1^\prime\sum_{i=1}^3\cos^4\theta_{0i}+\xi_2^\prime\sum_{i=1}^3\cos^2\theta_{0i}\right] \nonumber \\
& & -\frac{1}{3}\xi_3^\prime\left[\sin2\theta_{0j}\sum_{i=1}^3\sin2\theta_{0i}-\sum_{i=1}^3\sin^2 2\theta_{0i}\right] \nonumber \\
& = & 2\left[\xi_1^\prime(9/8)+\xi_2^\prime(3/2)\right]-(1/3)\xi_3^\prime\left[\sin2\theta_{0j}\times 0-(3/2)\right] \nonumber \\
& = & (9\xi_1^\prime+12\xi_2^\prime+2\xi_3^\prime)/4,
\end{eqnarray}
and
\begin{eqnarray}
\delta\theta_i & = & \frac{2}{3}\sum_{j=1}^3\cos^2\theta_{0j}\xi_3^\prime(\sin2\theta_{0j}-\sin2\theta_{0i}) \nonumber \\
& & -\sum_{j=1}^3\sin 2\theta_{0j}(\xi_1^\prime\cos^2\theta_{0j}+\xi_2^\prime) \nonumber \\
& = & \frac{2\xi_3^\prime}{3}\left[2\sum_{j=1}^3\cos^3\theta_{0j}\sin\theta_{0j}-\sin2\theta_{0i}\sum_{j=1}^3\cos^2\theta_{0j}\right]\nonumber  \\
& & -2\xi_1^\prime\sum_{j=1}^3\cos^3\theta_{0j}\sin\theta_{0j}-\xi_2^\prime\sum_{j=1}^3\sin2\theta_{0j} \nonumber \\
&  = & (2\xi_3^\prime/3)\left[2\times 0-\sin2\theta_{0i}\times(3/2)\right]-2\xi_1^\prime\times 0-\xi_2^\prime\times 0 \nonumber \\
& = & -\xi_3^\prime\sin2\theta_{0i},
\end{eqnarray}
respectively, leading to (\ref{xi_dependence}).

\section{Poisson's ratio}\label{Poisson}
In order to find the Poisson's Ratio $\nu$, ($\nu=-\varepsilon_\perp/\varepsilon$),
we need to find the transverse strain
$\varepsilon_\perp=\delta L_\perp/L_{\perp 0}$, where $L_{\perp 0}$ is a length of
the material perpendicular to the strain direction and $\delta L_{\perp}$ its
deformation upon tensile strain $\varepsilon$.
If $\mathbf{L}_{\perp 0}=t_a\mathbf{a}+t_b\mathbf{b}$ is a lattice vector,
which is perpendicular to the vector $\mathbf{L}_0=n\mathbf{a}+m\mathbf{b}$,
which defines the strain direction, then 
\begin{equation}
\begin{aligned}
& \mathbf{L}_{\perp 0}\mathbf{L}_0=0 \Rightarrow (t_a\mathbf{a}+t_b\mathbf{b})(n\mathbf{a}+m\mathbf{b})=0 \Rightarrow \\
& t_an(3a_0^2)+t_bm(3a_0^2)+(t_am+t_bn)(3a_0^2)/2=0 \Rightarrow \\
& t_a(2n+m)+t_b(2m+n)=0.
\end{aligned}
\end{equation}
For convenience we may select $t_a$ and $t_b$ to be $t_a=2m+n$ and 
$t_b=-(2n+m)$. Using (\ref{a_b}), $\mathbf{L}_{\perp 0}$ becomes
$\mathbf{L}_{\perp 0}=(m-n)\mathbf{r}_1+(2n+m)\mathbf{r}_2-(2m+n)\mathbf{r}_3$.
The projection of the deformation of a bond vector normal to the
strain direction is given by (\ref{y_deform}). Thus, the deformation
$\delta L_{\perp}$ of $\mathbf{L}_{\perp 0}$ is
\begin{equation}
\delta L_\perp = \sum_{i=1}^3 q_{\perp i}(\delta l_i\sin\theta_{0i}+a_0\cos\theta_{0i}\delta\theta_i),
\end{equation} 
where $q_{\perp 1}=m-n=2L_0/(3a_0)(\cos\theta_{03}-\cos\theta_{02})$,
$q_{\perp 2}=2n+m=(n+m)+n=2L_0/(3a_0)(\cos\theta_{01}-\cos\theta_{03})$ and
$q_{\perp 3}=-(2m+n)=-m-(n+m)=2L_0/(3a_0)(\cos\theta_{02}-\cos\theta_{01})$.
Using (\ref{sin_cos_cos}) we have
\begin{equation}
q_{\perp i}=2L_0/(\sqrt{3}a_0)\sin\theta_{0i},
\end{equation}
and consequently (using (\ref{l_lambda}), (\ref{lambda_i}), 
(\ref{xi_dependence_2}) and (\ref{sin2}))
\begin{eqnarray}
\delta L_{\perp} & = & \frac{2L_0}{\sqrt{3}a_0}\sum_{i=1}^3 \sin\theta_{0i}(\delta l_i\sin\theta_{0i}+a_0\cos\theta_{0i}\delta\theta_i) \nonumber \\ 
 & = & \frac{2L_0}{\sqrt{3}a_0}\sum_{i=1}^3 \left[\sin^2\theta_{0i}3a_0(\xi_1\cos^2\theta_{0i}+\xi_2) \right. \nonumber \\
& & \left. +a_0\sin\theta_{0i}\cos\theta_{0i}(-\xi_3\sin2\theta_{0i})\right]\varepsilon \nonumber \\
& = & 2\sqrt{3}L_0\left[\frac{\xi_1}{4}\sum_{i=1}^3 \sin^22\theta_{0i}+\xi_2\sum_{i=1}^3 \sin^2\theta_{0i} \right. \nonumber \\
& & \left. -\frac{\xi_3}{6} \sum_{i=1}^3\sin^22\theta_{0i} \right]\varepsilon   \nonumber \\
& = & (3/2)2\sqrt{3}L_0\left(\xi_1/4+\xi_2-\xi_3/6\right)\varepsilon.
\end{eqnarray}
The magnitude $L_{\perp 0}$ of the vector $\mathbf{L}_{\perp 0}$ is
\begin{eqnarray}
\L_{\perp 0} & = & |(2m+n)\mathbf{a}-(2n+m)\mathbf{b}| = 
|-q_{\perp 3}\mathbf{a}-q_{\perp 2}\mathbf{b}|\nonumber \\
& = & 2L_0/(\sqrt{3}a_0)|\sin\theta_{02}\mathbf{a}+\sin\theta_{03}\mathbf{b}| \nonumber \\
& = & 2L_0(\sin^2\theta_{02}+\sin^2\theta_{03}+\sin\theta_{02}\sin\theta_{03})^{1/2}. \nonumber
\end{eqnarray}
Using (\ref{sin1}) and (\ref{sin2})
\begin{equation}
\begin{aligned}
&\sin^2\theta_{02}+\sin^2\theta_{03}+\sin\theta_{02}\sin\theta_{03} \\
& = (1/2)(\sin^2\theta_{02}+\sin^2\theta_{03})+ \\
& \qquad (1/2)(\sin^2\theta_{02}+\sin^2\theta_{03}+2\sin\theta_{02}\sin\theta_{03}) \\
& = (1/2)(3/2-\sin^2\theta_{01})+(1/2)(\sin\theta_{02}+\sin\theta_{03})^2 \\
& = 3/4-(1/2)\sin^2\theta_{01}+(1/2)\sin^2\theta_{01} = 3/4. \nonumber
\end{aligned}
\end{equation}
Thus,
\begin{equation}
L_{\perp 0} = 2L_0\sqrt{3/4} = \sqrt{3}L_0,
\end{equation}
and consequently,
\begin{equation}
\varepsilon_\perp = \delta L_\perp/L_{\perp 0}=(3\xi_1/4+3\xi_2-\xi_3/2)\varepsilon,
\end{equation}
which leads to (\ref{nu}).

\section{Derivation of Eqs.~(\ref{relations1}), (\ref{relation1}),
(\ref{relation2}), (\ref{relation3})}\label{relations}

The first of (\ref{relations1}) can be directly obtained if we
divide by parts (\ref{xi_1}) and (\ref{xi_3}).
Using that equation, (\ref{xi_2}) becomes
\begin{eqnarray}\label{G1}
\xi_2 & = & k_{s1}k_{s2}(1-18k_b^\prime/k_{s1})/K_0 \nonumber \\
      & = & k_{s1}k_{s2}(1-(9/2)\xi_1/\xi_3)/K_0.
\end{eqnarray}
(\ref{xi_3}) can also be written as
\begin{equation}\label{G2}
\xi_3=2k_{s1}k_{s2}(6+k_{s1}/k_{s2})/K_0
\end{equation}
Dividing (\ref{G1}) and (\ref{G2}) by parts we obtain
\begin{equation}
\begin{aligned}
&  \xi_2/\xi_3 = \left[1-(9/2)\xi_1/\xi_2\right]/\left[2(6+k_{s1}/k_{s2})\right] \Rightarrow & \\
&  (6+k_{s1}/k_{s2})\xi_2 = \xi_3/2-9\xi_1/4 \Rightarrow &\\
&  k_{s1}/k_{s2} = (2\xi_3-9\xi_1)/(4\xi_2)-6 \Rightarrow & \\
&  k_{s2}/k_{s1} = 4\xi_2/(2\xi_3-9\xi_1-24\xi_2). &
\end{aligned}
\end{equation}
Using the first of (\ref{xi_dependence_2}), this equation leads to the second of
(\ref{relations1}).

From the expression $K_0=k_{s1}^2+9k_{s1}k_{s2}+18(k_{s1}+3k_{s2})k_b^\prime$,
it is obvious that the expression $k_{s1}k_{s2}+2(2k_{s1}+3k_{s2})k_b^\prime$,
which appears in (\ref{energy_gr2}), is
$k_{s1}k_{s2}+2(2k_{s1}+3k_{s2})k_b^\prime=(K_0-k_{s1}^2+18k_{s1}k_b^\prime)/9$.
Thus, using (\ref{G1}) and the second of (\ref{relations1}),
\begin{eqnarray}
& & (k_{s1}k_{s2}+2(2k_{s1}+3k_{s2})k_b^\prime)/K_0 \nonumber \\
& = & (1-k_{s1}(k_{s1}-18k_b^\prime)/K_0)/9 = \nonumber \\
& = & (1-(k_{s1}/k_{s2})(k_{s2}(k_{s1}-18k_b^\prime)/K_0))/9 = \nonumber \\
& = & (1-\xi_2(k_{s1}/k_{s2}))/9 \nonumber \\
& = & (1+\xi_2(1-\xi_3+3\xi_2)/\xi_2)/9 \nonumber \\
& = & (2-\xi_3+3\xi_2)/9. \nonumber
\end{eqnarray}
Thus,
\begin{eqnarray}
A & = & (k_{s1}+6k_{s2})(2-\xi_3+3\xi_2)/18 \nonumber \\
  & = & k_{s1}(1+6k_{s2}/k_{s1})(2-\xi_3+3\xi_2)/18 \nonumber \\
  & = & k_{s1}[1-6\xi_2/(1-\xi_3+3\xi_2)](2-\xi_3+3\xi_2)/18 \nonumber \\
  & = & k_{s1}[(1-\xi_3-3\xi_2)/(1-\xi_3+3\xi_2)](2-\xi_3+3\xi_2)/18. \nonumber
\end{eqnarray}
Using (\ref{xi_dependence_2}) (i.e. $2-\xi_3=9\xi_1/2+6\xi_2$), 
we find
\begin{eqnarray}
A & = & k_{s1}\left(\frac{1-\xi_3-3\xi_2}{1-\xi_3+3\xi_2}\right)\frac{9\xi_1/2+6\xi_2+3\xi_2}{18} \nonumber \\
  & = & k_{s1}\left(\frac{1-\xi_3-3\xi_2}{1-\xi_3+3\xi_2}\right)\frac{\xi_1+2\xi_2}{4}.
\end{eqnarray}
Solving this equation with respect to $k_{s1}$ we get (\ref{relation1}).

Using the expression in (\ref{relation1}) for $k_{s1}$ and (\ref{relations1}), 
the derivation of (\ref{relation2}) and (\ref{relation3}) is obvious.

\section{Useful relations between trigonometric functions of $\theta_{0i}$
of graphene}\label{useful}

Some relations, which are used in the present study, between the trigonometric
functions of the angles $\theta_{0i}$ defined by (\ref{theta_0i}), are
presented here. 

As we have already seen in Sec.~\ref{strain_constr}, $q_1=-m$, $q_2=-n$
and $q_3=n+m$. Thus, $q_1+q_2+q_3=0$, and consequently,
$\cos\theta_{01} + \cos\theta_{02} + \cos\theta_{03}=0$, where 
$\theta_{0i}=\theta_{0i}(\theta_0)=2\pi i/3-\theta_0$, $i=1,2,3$.
Obviously, (i) $2\theta_{01}(\theta_0)=4\pi/3-2\theta_0=\theta_{02}(2\theta_0)$,
$2\theta_{02}(\theta_0)=8\pi/3-2\theta_0=2\pi+\theta_{01}(2\theta_0)$ and
$2\theta_{03}(\theta_0)=4\pi-2\theta_0=2\pi+\theta_{03}(2\theta_0)$, and 
(ii) $4\theta_{01}(\theta_0)=8\pi/3-4\theta_0=2\pi+\theta_{01}(4\theta_0)$,
$4\theta_{02}(\theta_0)=16\pi/3-2\theta_0=4\pi+\theta_{02}(4\theta_0)$ and
$4\theta_{03}(\theta_0)=8\pi-2\theta_0=6\pi+\theta_{03}(4\theta_0)$.
Consequently, for $k=1$, or 2, or 4,
\begin{equation}\label{cos1}
\cos(k\theta_{01}) + \cos(k\theta_{02}) + \cos(k\theta_{03})=0.
\end{equation}
The first derivative of the above equation with respect to $\theta_0$ gives
\begin{equation}\label{sin1}
\sin(k\theta_{01}) + \sin(k\theta_{02}) + \sin(k\theta_{03})=0.
\end{equation}
Using (\ref{cos1}) for $k=2$ or $k=4$, and the relation $\cos 2\theta=2\cos^2\theta-1$ we obtain
\begin{eqnarray}\label{cos2}
\cos^2(2\theta_{01})+\cos^2(2\theta_{02})+\cos^2(2\theta_{03}) & = & \nonumber \\
\cos^2\theta_{01}+\cos^2\theta_{02}+\cos^2\theta_{03} & = & 3/2.
\end{eqnarray}
Then, using the relation $\sin^2\theta=1-\cos^2\theta$, we obtain
\begin{eqnarray}\label{sin2}
\sin^2(2\theta_{01})+\sin^2(2\theta_{02})+\sin^2(2\theta_{03}) & = & \nonumber \\
\sin^2\theta_{01}+\sin^2\theta_{02}+\sin^2\theta_{03} & = & 3/2.
\end{eqnarray}
Moreover, using the relation $\sin 2\theta=2\sin\theta\cos\theta$, (\ref{sin1})
for $k=2$ yields
\begin{equation}\label{sin_cos}
\sin\theta_{01}\cos\theta_{01}+\sin\theta_{02}\cos\theta_{02}+\sin\theta_{03}\cos\theta_{03}=0.
\end{equation}
Using (\ref{cos1}) for $k=1$ and (\ref{cos2}), we obtain
\begin{equation}
\begin{aligned}
& (\cos\theta_{01} + \cos\theta_{02} + \cos\theta_{03})^2 = 0  \Rightarrow  \\
& \cos^2\theta_{01}+\cos^2\theta_{02}+\cos^2\theta_{03}+ 2\cos\theta_{01}\cos\theta_{02} +   \\
& \qquad \qquad 2\cos\theta_{02}\cos\theta_{03} +2\cos\theta_{03}\cos\theta_{01} = 0  \Rightarrow  \\
& \cos\theta_{01}\cos\theta_{02}+\cos\theta_{02}\cos\theta_{03} \\
& \qquad \qquad \qquad +\cos\theta_{03}\cos\theta_{01}=-3/4.
\end{aligned}
\end{equation}
In turn, using (\ref{sin1}) for $k=1$ and (\ref{sin2}), we obtain
\begin{equation}
\sin\theta_{01}\sin\theta_{02}+\sin\theta_{02}\sin\theta_{03}+\sin\theta_{03}\sin\theta_{01} = -\frac{3}{4}.
\end{equation}
Thus, 
\begin{equation}\label{cos2cos2}
\begin{aligned}
& (\cos\theta_{01}\cos\theta_{02}+\cos\theta_{02}\cos\theta_{03}+ \\
& \qquad \qquad \cos\theta_{03}\cos\theta_{01})^2 = 9/16 \Rightarrow \\
& \cos^2\theta_{01}\cos^2\theta_{02}+\cos^2\theta_{02}\cos^2\theta_{03}+\cos^2\theta_{03}\cos^2\theta_{01}+  \\
& 2\cos\theta_{01}\cos\theta_{02}\cos\theta_{03}(\cos\theta_{01}+\cos\theta_{02}+ \\
& \qquad \qquad \cos\theta_{03}) = 9/16 \Rightarrow \\
& \cos^2\theta_{01}\cos^2\theta_{02}+\cos^2\theta_{02}\cos^2\theta_{03}+ \\
& \qquad \qquad \cos^2\theta_{03}\cos^2\theta_{01} = 9/16.
\end{aligned}
\end{equation}
Consequently, (\ref{cos2}) gives
\begin{equation}\label{cos4}
\begin{aligned}
& (\cos^2\theta_{01}+\cos^2\theta_{02}+\cos^2\theta_{03})^2=9/4 \Rightarrow \\
& \cos^4\theta_{01}+\cos^4\theta_{02}+\cos^4\theta_{03} + 2(\cos^2\theta_{01}\cos^2\theta_{02}+ \\
& \cos^2\theta_{02}\cos^2\theta_{03}+\cos^2\theta_{03}\cos^2\theta_{01})= 9/4 \Rightarrow  \\
& \cos^4\theta_{01}+\cos^4\theta_{02}+\cos^4\theta_{03} = 9/8,
\end{aligned}
\end{equation}
and in turn, (\ref{sin2}) gives
\begin{eqnarray}
& & (\sin^2\theta_{01}+\sin^2\theta_{02}+\sin^2\theta_{03})^2=9/4 \Rightarrow \nonumber \\
& & \sin^4\theta_{01}+\sin^4\theta_{02}+\sin^4\theta_{03} + 2(\sin^2\theta_{01}\sin^2\theta_{02}+ \nonumber \\
& & \sin^2\theta_{02}\sin^2\theta_{03}+\sin^2\theta_{03}\sin^2\theta_{01})= 9/4 \Rightarrow \nonumber \\
& & \sin^4\theta_{01}+\sin^4\theta_{02}+\sin^4\theta_{03} = 9/8.
\end{eqnarray}
Taking the first derivative of (\ref{cos4}), we obtain
\begin{equation}\label{cos3_sin1}
\sum_{i=1}^3 \cos^3\theta_{0i}\sin\theta_{0i} = 0.
\end{equation}

Moreover, let us consider the trigonometric identity
\begin{multline*}
 \sin(m\theta_{0j})-\sin(m\theta_{0i}) = \\
 =2\sin[m(\theta_{0j}-\theta_{0i})/2]\cos[m(\theta_{0i}+\theta_{0j})/2].
\end{multline*}
For the angles $\theta_{0i}$ in (\ref{theta_0i}) we have 
$(\theta_{0j}-\theta_{0i})/2=(j-i)\pi/3$
and $(\theta_{0j}+\theta_{0i})/2=(i+j)\pi/3-\theta_0$.
For $(i,j,k)=(1,2,3)$, or $(2,3,1)$, or $(3,1,2)$, the sum
$i+j+k=6$, and consequently, $i+j=6-k$.
Thus, $(i+j)\pi/3-\theta_0=(6-k)\pi/3-\theta_0=2\pi-k\pi/3-\theta_0=
2\pi-k\pi+\theta_{0k}$. Consequently, 
$\cos[m(\theta_{0i}+\theta_{0j})/2]=\cos[m(\theta_{0k}-k\pi)]$ and
$\sin[m(\theta_{0j}-\theta_{0i})/2]=\sin[m(j-i)\pi/3]$. Thus,
for $m=1$
\begin{equation}\label{cos_sin_sin_1}
\sin\theta_{0j^\prime}-\sin\theta_{0i^\prime} = -\sqrt{3}\cos\theta_{0k^\prime},
\end{equation}
and for $m=2$
\begin{equation}\label{cos_sin_sin_2}
\sin2\theta_{0j^\prime}-\sin2\theta_{0i^\prime} = \sqrt{3}\cos2\theta_{0k^\prime}.
\end{equation}
Taking the first derivative of (\ref{cos_sin_sin_1}) with respect to
$\theta_0$ we obtain
\begin{equation}\label{sin_cos_cos}
\cos\theta_{0j^\prime}-\cos\theta_{0i^\prime} = \sqrt{3}\sin\theta_{0k^\prime}.
\end{equation}



\end{document}